\documentclass{aa}
\usepackage{txfonts} 
\usepackage[OT1]{fontenc}
\usepackage[latin1]{inputenc} 
\usepackage{longtable} 
\usepackage{supertabular}
\usepackage{graphicx}

\def\mathrelfun#1#2{\lower3.6pt\vbox{\baselineskip0pt\lineskip.9pt
  \ialign{$\mathsurround=0pt#1\hfil##\hfil$\crcr#2\crcr\sim\crcr}}}

\def\fun#1#2{\lower3.6pt\vbox{\baselineskip0pt\lineskip.9pt
  \ialign{$\mathsurround=0pt#1\hfil##\hfil$\crcr#2\crcr\sim\crcr}}}
\def\etal{{et~al. }}   

\begin{document}
\title{Submillimetre point sources from the Archeops experiment:\\
  Very Cold Clumps in the Galactic Plane } \titlerunning{{\sc
    Archeops} Submillimetre Point Sources} 
\author{
  F.--X.~Désert~\inst{1} \and J.~F.~Mac\'{\i}as--P\'erez ~\inst{2}
  \and F.~Mayet~\inst{2} \and G.~Giardino~\inst{7} \and
  C.~Renault~\inst{2} \and J.~Aumont~\inst{2} \and
  A.~Beno\^{\i}t~\inst{3} \and J.--Ph.~Bernard~\inst{4} \and
  N.~Ponthieu~\inst{5} \and M.~Tristram~\inst{6} }

   \offprints{desert@obs.ujf-grenoble.fr}

\institute{
Laboratoire d'Astrophysique, Obs. de Grenoble, BP 53,
38041 Grenoble Cedex 9, France
\and
LPSC, Universit\'e Joseph Fourier Grenoble 1, CNRS/IN2P3, Institut National
Polytechnique de Grenoble, 53 avenue des Martyrs, 38026 Grenoble cedex, France \and
Centre de Recherche sur les Très Basses Températures,
BP166, 38042 Grenoble Cedex 9, France
\and
Centre d'Étude Spatiale des Rayonnements,
BP 4346, 31028 Toulouse Cedex 4, France
\and
Institut d'Astrophysique Spatiale, Bât. 121, Université Paris~XI,
91405 Orsay Cedex, France \and
Laboratoire de l'Accélérateur Linéaire, BP 34, Campus
Orsay, 91898 Orsay Cedex, France \and
ESA - Research and Science Support Department, ESTEC, Postbus 299, 2200 AG Noordwijk                    
The Netherlands
}

  \date{\today}

  \abstract{} {Archeops is a balloon--borne experiment, mainly
    designed to measure the Cosmic Microwave Background (CMB)
    temperature anisotropies at high angular resolution ($\sim 12$
    arcminutes). By--products of the mission are shallow sensitivity
    maps over a large fraction of the sky (about 30~\%) in the
    millimetre and submillimetre range at 143, 217, 353 and 545~GHz.
    From these maps, we produce a catalog of bright submillimetre
    point sources.}  {We present in this paper the processing and
    analysis of the Archeops point sources.  Redundancy across
    detectors is the key factor allowing us to distinguish glitches from
    genuine point sources in the 20 independent maps.}  {We look at
    the properties of the most reliable point sources, totalling 304.
    Fluxes range from 1 to 10,000 Jy (at the frequencies covering 143
    to 545~GHz). All sources are either planets (2) or of galactic
    origin.  The longitude range is from 75 to 198~degrees. Some of the
    sources are associated with the well-known Lynds Nebulae and HII
    compact regions in the galactic plane. A large fraction of the
    sources have an IRAS counterpart.  Except for Jupiter, Saturn, the
    Crab and Cas~A, all sources show a dust-emission--like modified
    blackbody emission spectrum.  Temperatures cover a range from 7 to
    27~K. For the coldest sources ($T<10\,\mathrm{K}$), a steep
    $\nu^{\,\beta}$ emissivity law is found with a surprising
    $\beta\sim\,3\,\,\mathrm{to}\,\,4$. An inverse relationship
    between $T$ and $\beta$ is observed. The number density of sources
    at 353~GHz with flux brighter than 100~Jy is of the order of 1 per
    degree of Galactic longitude.  These sources will provide a strong
    check for the calibration of the Planck HFI focal plane geometry
    as a complement to planets. These very cold sources observed by Archeops
    should be prime targets for mapping observations by the Akari and
    Herschel space missions and ground--based observatories. } {}

\maketitle


\section{Introduction}

{\sc Archeops} is a balloon--borne experiment following on from the
{\sc Planck} satellite and its High Frequency Instrument ({\sc HFI}).
It measures the Cosmic Microwave Background (CMB) temperature
anisotropies at high angular resolution ($\sim 12$ arcminutes) over a
large fraction of the sky (around 30~\%) in the millimetre and
submillimetre range at 143, 217, 353 and 545~GHz. The main results of
cosmological nature have been discussed elsewhere
(\cite{archpaper,archpaper_cospar,tristram_cl}). But because we have,
for the first time, a large survey of the millimetre sky, studies on
other scientific topics can be performed.  Detection of large--scale
polarized dust emission is reported in \cite{archpolar, ponthieu05}.
The large scale spectral properties of the dust emission have been
investigated by \cite{jpbcospar} and finally, the statistical
detection of clusters of galaxies is shown by \cite{carlos_sz}.

Here, we analyze another by--product of the {\sc Archeops} mission.
We look at the properties of the most reliable point sources in the
{\sc Archeops} survey. We discuss the extraction method, the catalog
of candidate cold dust clumps of likely galactic origin, as well as
two planets and two supernova remnants. Some of these clumps are
producing massive stars.  Implications for the galactic clump mass
distribution function and the expected number of Galactic point
sources in the {\sc Planck} survey are then discussed.

\section{Point--source extraction}

\subsection{The instrument}

The gondola contains a primary mirror with an effective 1.5~m
diameter, a secondary mirror and a photometer containing spider--web
bolometers cooled to 100~mK. The instrument is described in detail by
\cite{trapani} and \cite{processing}. The {\sc Archeops}
353~GHz channel consists of three pairs of bolometers mounted on
polarizer dichroics so as to detect the polarized diffuse emission of
Galactic dust.  The telescope boresight angle is 48~deg. with respect
to the zenith. The gondola is made spins at 2~rpm via a motor fixed
on the balloon chain. The Eastward balloon trajectory and the Earth
rotation make the instantaneous circle  drift on the celestial
sphere. This scanning strategy produces a shallow survey of a large
fraction of the sky in few hours. The angular resolution varies from
15 to 8 arcminutes (FWHM) with the channel frequency from 143 to
545~GHz. The sensitivities and main data processing methods are
described by \cite{processing}. We use the data from 6 detectors at
143~GHz, 7 at 217~GHz, 6 at 353~GHz and one at 545~GHz.

\subsection{The method}

Because of the scanning strategy, one detector will sweep rapidly
a given diffraction spot of the sky. Instead of using the timeline
signature of point sources (impulse convolved with beam response)
which can be confused with glitches (impulse only), we prefer to use
the redundancy between the 20 available bolometers at the map level.
Glitches will fall at random locations, whereas point sources will
produce a concentration of bright spots in the same sky position in
several bolometer maps. The detection process is a separate step from
the measurement process. Once a candidate location is found, the
point--source flux and error are measured on all bolometer maps and
coadded with natural weights.

The data from the last flight (KS3, 2002, Feb. 7th) only are used,
with a time range between 15.4 and 27.3 UT (well within night time).
What is projected on sky maps is between 15.5 and 27~UT.

\subsection{Timeline processing}

\begin{figure}[htbp]
\begin{center}
 \includegraphics[%
  width=0.8\columnwidth,
  keepaspectratio,
  angle=90]{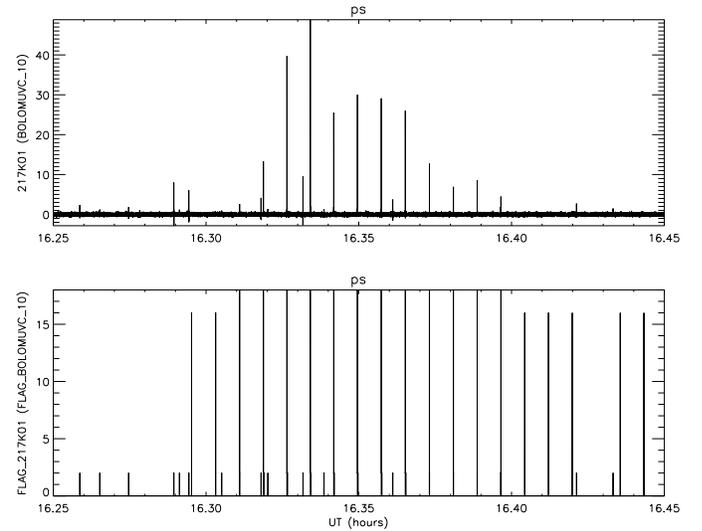}
\end{center}

\caption{\label{fig:clean3}Extract of the bolometer 217K01 signal
  timeline (upper plot). Note the regular spikes produced by Jupiter,
  when the gondola crosses its line of sight at each revolution
  (pendulation explains why the spike intensity is not smooth during
  the different crossings).  On the other hand, cosmic ray hits happen
  at random times. The lower plot indicates the cumulated flagging
  found during data processing (a value of 2 for glitches and a value
  of 16 around known point sources).}
\end{figure}

The Galactic timeline processing is described by \cite{processing} and
is slightly different from the CMB one. An example is shown in
Fig.~\ref{fig:clean3}. The low frequency thermal drifts, the
atmospheric emission and the Galactic diffuse emission signal produce
a varying background signal. Stripes can be produced that degrade the
efficiency of point-source detection.  Therefore, it is necessary to
subtract the baseline. The timeline, which is masked and interpolated
around glitches or strong point sources, is smoothed in order to
provide this baseline, to be subtracted from the initial time ordered
information (hereafter TOI).  The smoothing occurs, first over a
period of several revolutions, then over lengths of ninety degrees on
the instantaneous circle described by the gondola. Atmospheric noise
can leave residues on ten degree scales. A timeline component
separation is done on these scales or larger. The clean timeline
(CToi) is ready for map making. We also produce a clean deglitched
timeline (CDToi) to be used for faint point source and background map
computations.

\subsection{Map--making}

\begin{figure}[htbp]
\begin{center}
\includegraphics[%
  width=0.20\paperwidth,
  keepaspectratio,
  angle=90]{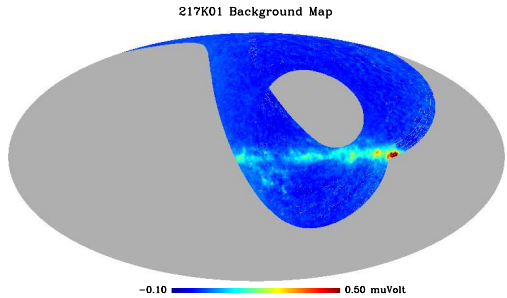}
\end{center}
\caption{\label{fig:mapps} Background map of the 217K01 detector.
  This map is obtained by projecting the clean deglitched timeline and
  by smoothing the result with a $60\,\mathrm{arcmin.}$ circular
  Gaussian kernel. An all-sky Mollweide projection, with the Galactic
  Anticenter at the center of the projection, is used throughout this
  paper. About 30 percent of the sky was observed. This smooth map is
  subtracted from the detector map in order to find the point sources
  by a simple thresholding algorithm.}
\end{figure}

The prepared timelines have white noise properties. Thus we proceed
with a simple map-making with a natural weighting of the prepared
timelines (CToi and CDToi) for each detector, using pointing
information from the star sensor and the known detector position in
the focal plane. The map is made in the Healpix scheme
(\cite{healpix}) with $N_{\mathrm{side}}=512$. It provides a 7--arcmin
pixel size which is adequate to sample the point-spread-function. The
map is smoothed with a 12--arcminute circular Gaussian kernel in
order to prepare for the point--source detection processing.  A
background component is evident with angular scales of a few degrees,
due to the Galaxy diffuse component emission (mostly due to
interstellar dust at these frequencies). We compute a smooth
background component (Fig.~\ref{fig:mapps}) by projecting the CDToi
and then smoothing the resulting map with a 60~arcminute circular
Gaussian kernel. The smoothing is done in the spherical harmonic
space. Pixels that were unobserved (at the edge of the survey) are
replaced by zero before the spherical harmonic transform is
done. The final background--subtracted map is then computed for each
detector.  Thus, the end map represents for each bolometer the sky
convolved with a 12--arcmin. Gaussian kernel minus a 60--arcmin.
Gaussian kernel. This spatial filter has a zero mean and is
appropriate for a simple point--source detection-by-thresholding
algorithm.  We also produce channel maps obtained by optimal
coaddition of those previous maps for detectors of a given channel
({\sl i.e.} at the same frequency). A relative calibration between
detectors is used for that coaddition, which is described in
\cite{processing}.

\subsection{Point-source extraction}
\label{psextraction}
Outside the galactic plane, the detector map is dominated by noise. A
Gaussian function is fit to the histogram of the product of the pixel
value $v_p$ by the square root of the number of hits $N_p$. For a
given detector, the dispersion $\sigma_d$ gives the typical elementary
noise value for one hit, {\sl i.e.} for one data sample, the data
acquisition rate being 153~Hz.  It is then used to estimate the
noise on each pixel as $\sigma_p=\sigma_d/\sqrt{N_p}$. A listing of
target pixels defined as $\frac{S}{N}=|\frac{v_p}{\sigma_p}| > 5$ is
produced for each detector. At this stage glitches and point sources
are not separated. Two catalogs are obtained depending on whether one
uses the CToi (still containing hits from strong point sources and
glitches) or the CDToi (having strong point sources and glitches
removed). For the CToi catalog, we perform a final separation of point
sources from glitches by requesting that, for a point--source, the
above 5~$\sigma$ detection criteria for a given sky pixel be matched by
at least 5 different detectors.  For the CDToi catalog, we use the
final point--source criterium that at least two channel maps have
4~$\sigma$ detections.  A channel map is defined here as the optimal
average of the maps of all detectors at the same frequency. Out of the
4500 pixels that satisfy the criteria, some are connected to each
other and correspond to the same point--source. To have a single
position for each point-source, we keep the pixel for which a maximum
number of detectors have a 5~$\sigma$ detection.  This defines a final
catalog of sources. For each of them, we can measure the flux, error
and position on the 20 independent detector maps and the 4 channel
maps.

\begin{figure*}[htbp]
\begin{center}
\includegraphics[%
   width=0.6\paperwidth,
   keepaspectratio,
  angle=90]{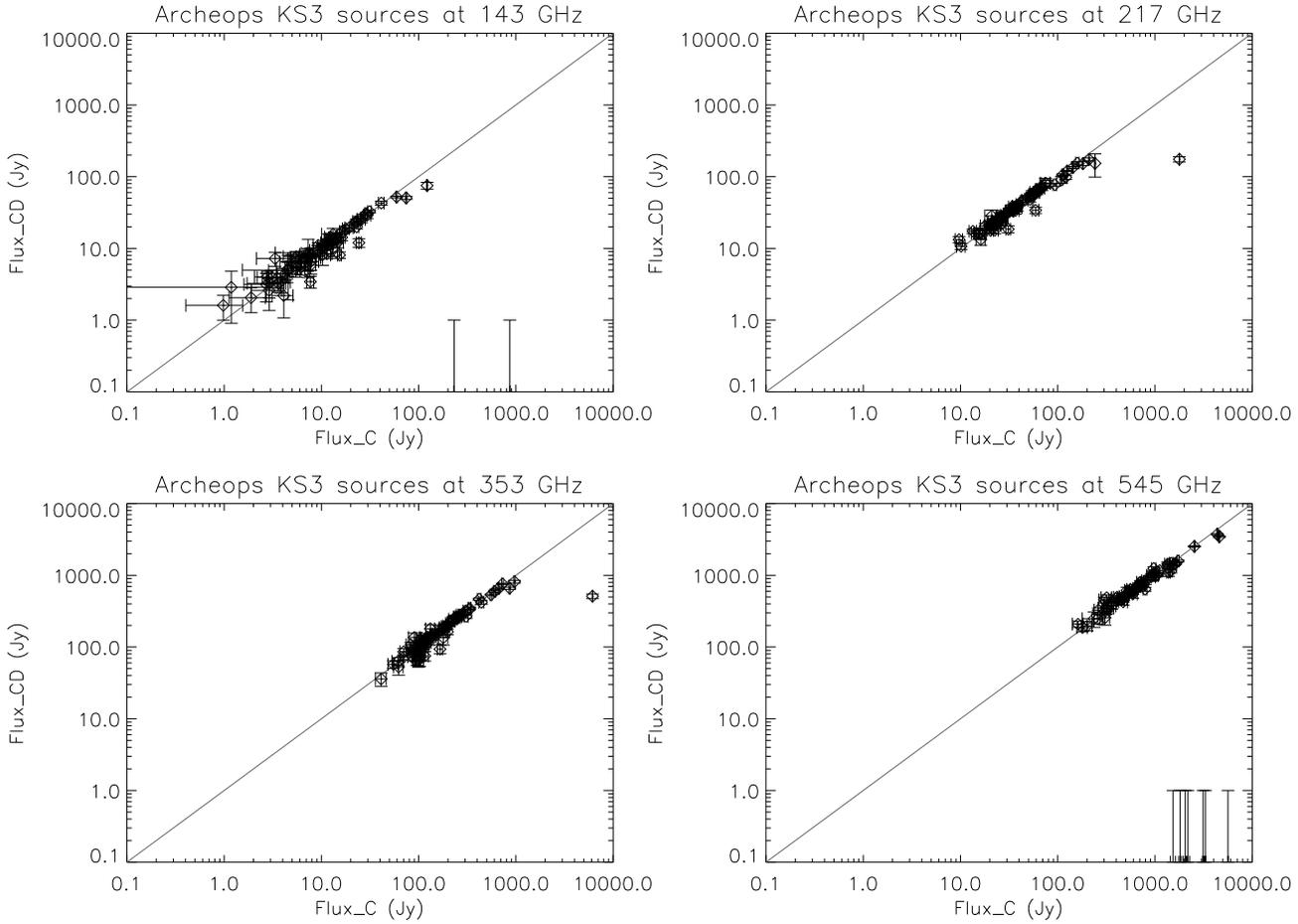}
\end{center}
\caption{\label{fig:glnogl}Flux comparison between the two detection
  methods at the four frequencies, for sources in common to both
  methods. The plots compare the flux obtained with (CD) and without
  (C) a deglitching applied to the data. }
\end{figure*}

Fig.~\ref{fig:glnogl} shows the flux comparison of the two methods
(CToi and CDToi) for the 87 sources in common in the two catalogs. A
very good agreement is obtained for most of the flux range and for all
four frequencies. At the bright end, we expect and observe that the
deglitching slightly biases the flux measured from the CDToi (``CD''
fluxes in the abscissae of the figure). However, the CDToi method is
more powerful at the faint end, because the signal to noise ratio
benefits from the coaddition. We have therefore merged the two
catalogs. For each of these candidates, we average the flux from
different bolometers in each of the 4 frequency bands with a natural
weight (equal to the inverse square of the noise). Positions are
measured with the same weighting. From the internal dispersion between
channel positions, we have estimated the $1\,\sigma$ position accuracy
to be about 4~arcmin (a third of the beam width). For 23 of the
brightest point sources, one of the channel position disagrees with
the final position. Nearby glitches in the CToi objects and confusion
might cause such a systematic effect.

\subsection{Flux calibration and planet observations}\label{Fluxcalib}

The flux calibration is performed at this stage. The CMB dipole
calibration is used at 143 and 217~GHz and the FIRAS galactic
calibration is used at 353 and 545~GHz (see details in
\cite{processing}). These are extended source calibrations. In order
to propagate them to point-source calibration, we integrate the beams
measured on Jupiter and Saturn. Assuming angular diameters for these
(unresolved) sources of 44.60 by 41.71 arcseconds and 19.05 by 17.00
arcseconds respectively, we obtain the brightness temperature given in
Table~\ref{tab:jupsat}. Statistical error bars are negligible. The
total error bars are made up from an absolute calibration error (resp.
4, 8, 12, and 8~\%), the estimated error due to the beam integration
procedure (estimated by comparing the beam shape obtained from Jupiter
and Saturn observations at the same frequency) and an intercalibration
error (the measurement dispersion across the bolometers at the same
frequency).

\begin{table}[th]
\begin{center}
\begin{tabular}{|c|c|c||c|c|} \hline
& \multicolumn{2}{c|}{{\bf Jupiter}} &\multicolumn{2}{c|}{{\bf Saturn}} \\ \hline \hline
$\rm \nu \ (GHz)$ & $\rm T_{RJ} \ (K)$  &  $\rm \sigma  \ (K)$        &
                    $\rm T_{RJ} \ (K)$  &  $\rm \sigma  \ (K)$\\ \hline
{\bf 143} & 165    & 20         & 160 & 19 \\ \hline
{\bf 217} & 139    & 22         & 144 & 23 \\ \hline
{\bf 353} & 159    & 24         & 179 & 27 \\ \hline
{\bf 545 }& 146    & 20         & 166 & 23 \\ \hline
\end{tabular}
\caption{ Jupiter and Saturn brightness temperature, as calibrated
  with Archeops extended source calibrators (the CMB dipole at 143 and
  217~GHz and the Firas Galaxy at 353 and 545~GHz). Error bars are
  absolute calibration errors.}
\label{tab:jupsat}
\end{center}
\end{table}

\begin{figure}[htbp]
\begin{center}
\includegraphics[%
   width=0.8\columnwidth,
   keepaspectratio,
  angle=0]{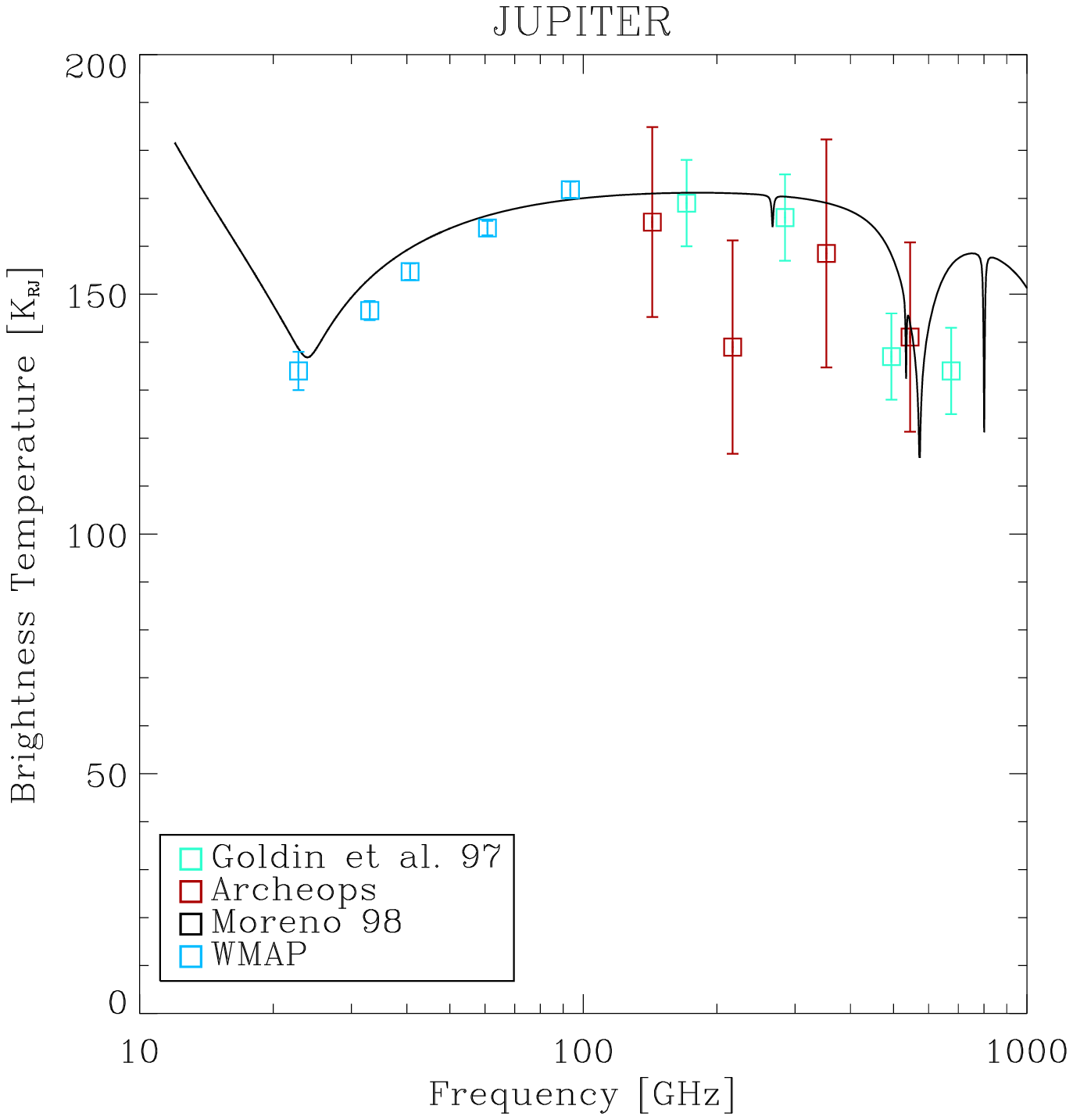}
\includegraphics[%
   width=0.8\columnwidth,
   keepaspectratio,
  angle=0]{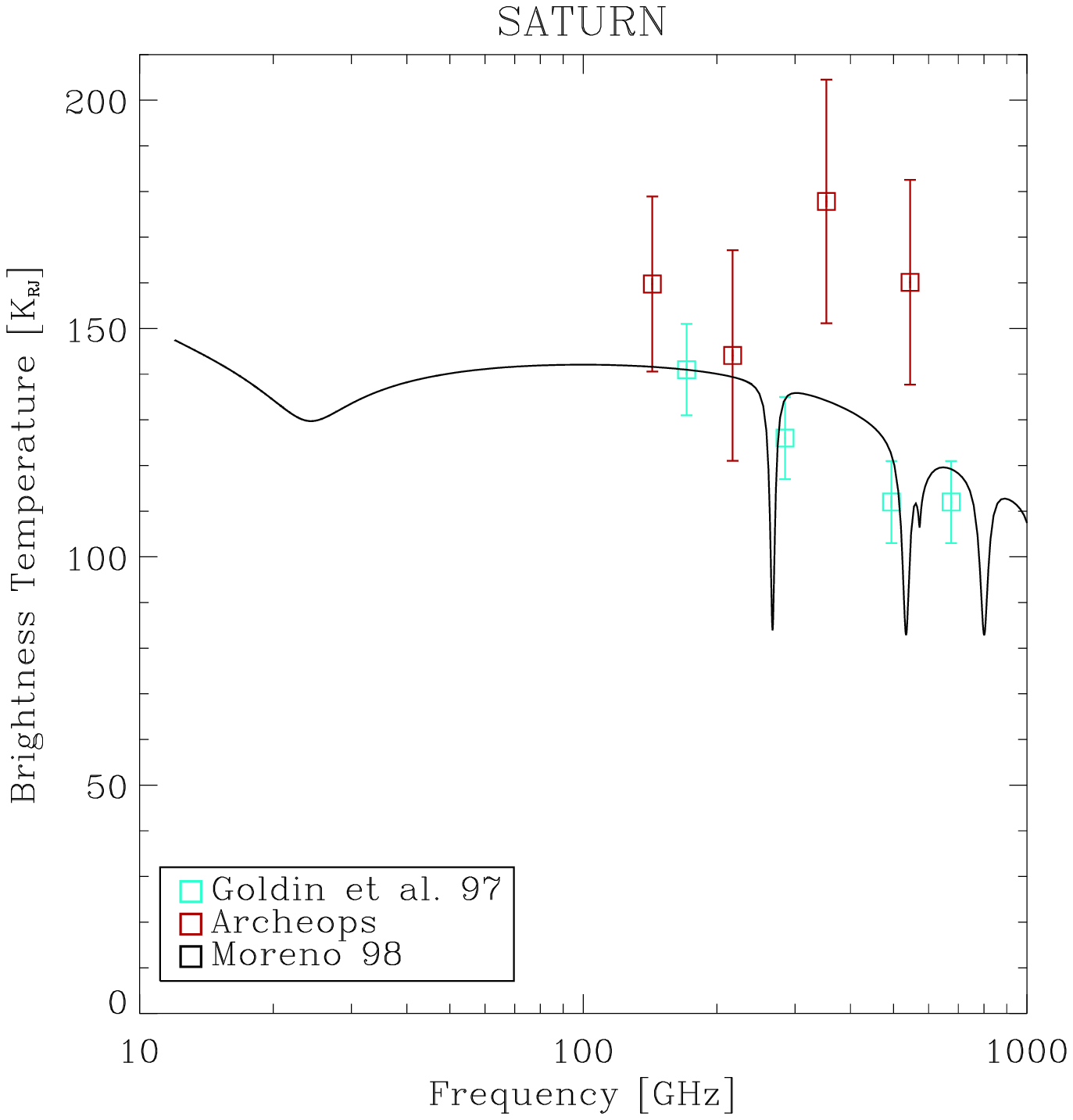}
\end{center}
\caption{\label{fig:JupSat} Jupiter (upper) and Saturn (lower) flux
  measurements (squares) compared with Moreno's atmosphere models
  (lines) (\cite{Moreno:1998}) as well as WMAP and
  \cite{goldin:1997}'s measurements.  }
\end{figure}

Figure~\ref{fig:JupSat} shows the (sub)millimetre spectrum of Jupiter
and Saturn, as measured by WMAP (\cite{Page2003}), {\sc Archeops} and
\cite{goldin:1997}.  There is a broad agreement of the flux scale over
a factor of 20 in frequency range. The ratio of Jupiter to Saturn flux
at a given frequency, which should be independent of the absolute flux
scale calibration, is larger in {\sc Archeops} than in the
\cite{goldin:1997} measurements.  Explanations might be sought in
small non-linearity problems in the {\sc Archeops} instrument for
these very high flux sources and also from the simplified emission
assumptions of the two planets (inclination of Saturn rings), as shown
by the detailed {\sc Pronaos} calibration (\cite{Pajot2006}).

In addition to the systematic errors discussed above, for the point
source catalog, we need to consider the error introduced by the flux
measurement method described in Sect.~\ref{psextraction}. The
convolution by a simplified 2D Gaussian kernel increases the
measurement intrinsic dispersion of bolometers at a single frequency.
For the final catalog, we estimate the absolute calibration scale
uncertainties as 14, 21, 17, and 15~\% at 143, 217, 353, and 545~GHz
respectively.

\section{Results}

\subsection{The point--source catalog}
\label{sect:pointsourcecatalog}

\begin{figure*}[htbp]
\begin{center}
\includegraphics[%
  width=0.5\textwidth,
  keepaspectratio,
  angle=270]{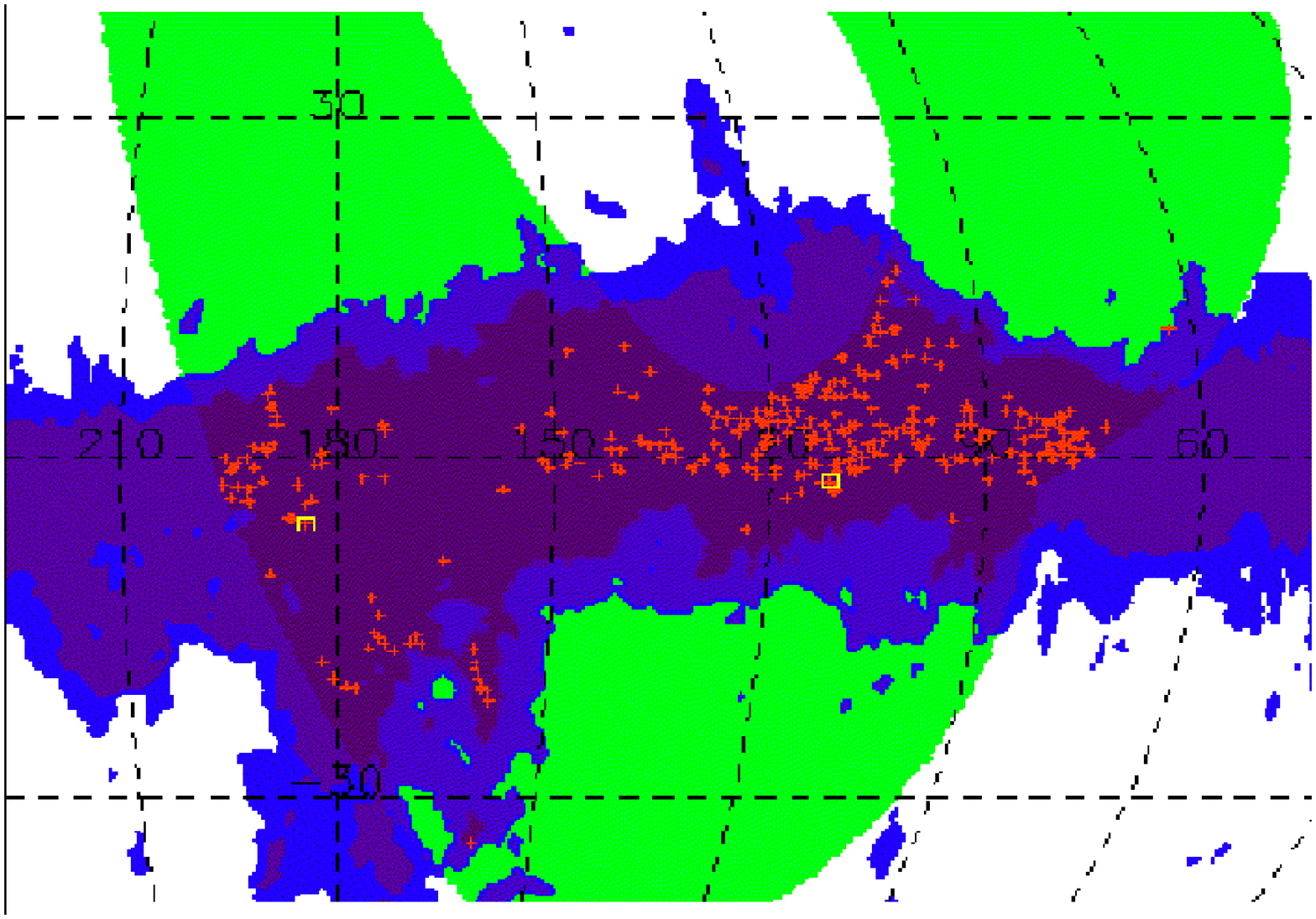}
\end{center}
\caption{\label{fig:map of detections} Location of the point sources
  (in red pluses) in galactic coordinates in a Mollweide
  projection. The Crab (Tau~A) and Cas~A correspond to the yellow
  squares (from left to right). Most of the sources are concentrated
  in the Galactic plane. The {\sc Archeops} survey is shown as the
  green area. A thresholded {\sc Iras} galactic map (blue) aids to
  trace high far--infrared background regions.}
\end{figure*}

{\sc Archeops} has detected 304 submillimetre point--sources in the
covered 30.0~\% fraction of the sky.  These are plotted in
Fig.~\ref{fig:map of detections}.  They are mostly in the Galactic
Plane, with a high concentration in the Cygnus and Taurus complexes.
Fluxes range from 1 to 10$^4$ Jy and median average fluxes and errors
are 3.0 and 1.4~Jy at 143~GHz, 19 and 3~Jy at 217~GHz, 76 and 13~Jy at
353~GHz, and 344 and 44~Jy at 545~GHz.  The survey is inhomogeneous in
sensitivity because the scanning strategy and the limited observing
time did not allow an equal number of pixel hits everywhere.  The
obtained median sensitivities are in agreement with expectations from
Table~8 by \cite{processing}, giving us some confidence in the
processing efficiency.  The average integration time spent by an {\sc
  Archeops} detector on a 12-arcmin spot on the sky is only
0.11~second.

The catalog of sources (the 2 planets being excluded) is given in
Table~\ref{catalog}. It includes the position on the sky with galactic
coordinates in degrees, fluxes in Jy, with their error bars and signal
to noise ratio. Associations with the {\sc Iras} point-source and
small extended source catalog are given in Table~\ref{assoc}. The
matching radius is 10 arcmin ($2.5\,\sigma$ of {\sc Archeops} position
accuracy). In many cases, there is more than one {\sc Iras}
counterpart within the matching radius, so the source with the
strongest flux at $100\,\mu{\mathrm{m}}$ ({\sc Iras}) was selected.
Associations with previously known sources are made with CDS
catalogs\footnote{{\tt http://cdsweb.u-strasbg.fr}}
(Table~\ref{assoc}).  Many sources have a counterpart with a bright or
dark Lynds Nebula or an HII compact (Sharpless) region (for example,
DR~21, W~3). Matches to HII regions with an angular size lower than 10
arcminutes from the catalogue of \cite{Paladini} are also included.
Finally, in Table~\ref{assoc}, sources whose {\sc Iras} counterpart is
likely an ultra-compact HII region (UCHII) are indicated. These were
identified from their {\sc Iras} color following the color criteria
from \cite{Kurtz} and applying a flux threshold of 100 Jy at
$100\,\mu{\mathrm{m}}$ (i.e. only sources with the colors of an
ultra-compact HII region and $F(100\,\mu{\mathrm{m}}) > 100$ Jy are
indicated).

\begin{figure}[htbp]
\begin{center}
\includegraphics[%
  width=0.6\columnwidth,
  keepaspectratio,
  angle=90]{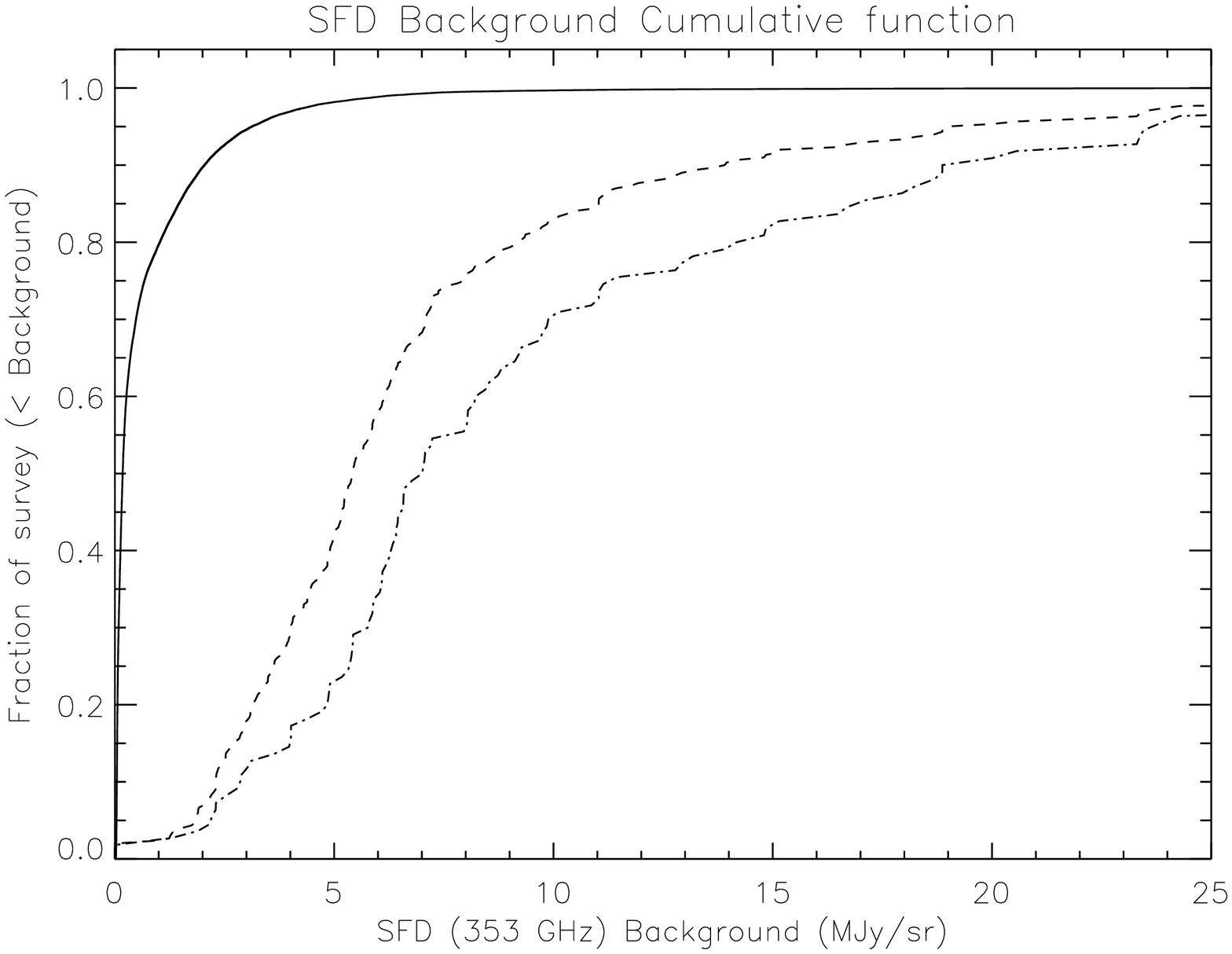}
\end{center}
\caption{Cumulated fraction (dashed curve) of sources falling below a
  given background. Here, sources are counted at the 353~GHz
  frequency. The background is measured as pixel value of a 353~GHz
  SFD map, smoothed with a 30~arcmin wide kernel. The dot-dash line
  shows the cumulated fraction of {\sc Archeops} point sources with a
  flux larger than 100~Jy.  The black curve shows the cumulated
  fraction of all pixels visited by {\sc Archeops}.
\label{fig:Background}}
\end{figure}

Concentration of the sources in the Galactic Plane and in molecular
cloud regions indicates a galactic origin.  This galactic origin of
most of the sources is confirmed in Fig.~\ref{fig:Background}. We
compute the background expected from {\sc Iras} extrapolations
(\cite{finkbeiner}) at the same frequency as the {\sc Archeops}
sources. There is not a direct correlation between the {\sc Archeops}
flux at 353~GHz and the diffuse background emission expected at the
same frequency, but we find in Fig.~\ref{fig:Background} that the
sources tend to gather in high background regions. For example, 60~\%
of the point sources are in regions with an average brightness larger
than 5~MJy/sr at 353~GHz.  Similarly, using the large scale Galactic
CO survey by \cite{Dame}, we do not find a direct correlation between
the CO velocity--integrated brightness and the {\sc Archeops} flux but
there is a strong trend for the submillimetre sources to lie in high
CO background regions. For example, half the point sources lie on top
of a CO background greater than $15\,\mathrm{K.km.s^{-1}}$ whereas half
of the pixels recorded by {\sc Archeops} with a CO measurement have a
CO flux greater than only $2.5\,\mathrm{K.km.s^{-1}}$

\subsection{Photometry}

We have compared the fluxes measured with {\sc Archeops} with previous
surveys conducted in the submillimetre domain.

In particular the Crab and Cas~A point--sources are well-known sources
and can be used for comparison. No other supernova remnant
(\cite{Green}) could be associated with the present catalog.

Concerning the Cas~A supernova remnant, {\sc Archeops} confirms the
submillimetre emission (Table~\ref{tab:cascrab}) in excess of the
synchrotron component and discovered and mapped by SCUBA
(\cite{Dunne2003}).  The 143~GHz point closely fits the synchrotron
extrapolation (giving some confidence in the {\sc Archeops}
photometric calibration), whereas a large excess exists at 217, 353,
and 545~GHz. The origin of this excess (cold dust, iron needles) has
been debated by \cite{Dwek2004} and \cite{Gomez2005} and is studied by
\cite{casa_paper}.

The Crab photometry is further analyzed in a companion paper
(\cite{crab_paper}). The 143~GHz measurement is in agreement with the
expected radio synchrotron component described by a power law with a
spectral index $\beta \simeq -0.299 \pm 0.009$ (\cite{Baars1977}).

\begin{table}[th]
\begin{center}
\begin{tabular}{|c|c|c||c|c|} \hline
                  & \multicolumn{2}{c|}{{\bf Cas A}} &\multicolumn{2}{c|}{{\bf
		  Crab}} \\ \hline \hline
$\rm \nu \ (GHz)$ & $\rm F \ (Jy)$  &  $\rm \sigma  \ (Jy)$        &
  $\rm F \ (Jy)$  &  $\rm \sigma   \ (Jy)$\\ \hline
{\bf 143} &  74.5    &  1.3           & 231.4 &   1.4 \\ \hline
{\bf 217} &  58.3    &  2.9           & 181.9 &   1.8 \\ \hline
{\bf 353} & 121      & 14             & 185   &  11   \\ \hline
{\bf 545} & 359      & 36             & 236   &  58   \\ \hline
\end{tabular}
\caption{Submillimetre fluxes and the statistical flux errors of the
  two supernova remnants detected by {\sc Archeops}.}
\label{tab:cascrab}
\end{center}
\end{table}

There is not a single common source with the WMAP catalog
(\cite{Hinshaw}), because the WMAP catalog contains only extragalactic
radio sources.

Sources observed by large ground--based telescopes are much fainter
than detected as a point--source by the {\sc Archeops} balloon
experiment.  We find that {\sc Archeops} fluxes are usually greater
than some of the ground--based measurements on integrated mapped
regions. However, most of the present sources have never been measured
at these frequencies and spatial resolution before. We can compare the
flux values with the few available values ($\sim 10$) published by
\cite{chini84,chinia,chinib}, and measured with a 90 arcsecond beam
ground-based photometer at 230~GHz. {\sc Archeops} values are always
above the ground-based observations by a factor that can reach 10. We
think that the photometric disagreement is due to the chopping
techniques used in ground-based experiments and to the spatial extent
of the sources.

No extragalactic point--source can be identified \footnote{We measured
  the Andromeda (M~31) extended source flux within {\sc Iras}
  100~$\mu$m contour of 5~MJy/sr (about one degree size), and found a
  marginal detection ($3\,\sigma$) at 545~GHz only: Fluxes of M31 :
  4407~Jy at {\sc Iras} 100~$\mu$m, $<26\,\mathrm{Jy}$ at 143~GHz,
  $<90\,\mathrm{Jy}$ at 217~GHz, $<340\,\mathrm{Jy}$ at 353~GHz and
  $660\pm200\,\mathrm{Jy}$ at 545~GHz. Upper limits are quoted as
  $2\,\sigma$. } in the present catalog.  The strongest known
extragalactic point source (for a 12-arcmin beam) is M~82 which has a
flux of about 5~Jy at 353~GHz.  It is not in our survey coverage, and
photometrically, it is below our sensitivity limit.

\subsection{Submillimetre spectra}\label{ss:submm}

Beside the 4 brightest sources (two planets and two SN remnants) most
of the sources have a steep spectrum rising in the submillimetre
domain with frequency, with typically an increase in flux by a factor
of 100 when going from 143~GHz to 545~GHz. That excludes synchrotron
or free--free emission as the main emission mechanism. Interstellar
dust is most likely the source of such spectra.

\begin{figure*}[htbp]
\begin{center}
\includegraphics[%
   width=0.7\paperwidth,
   keepaspectratio,
  angle=0]{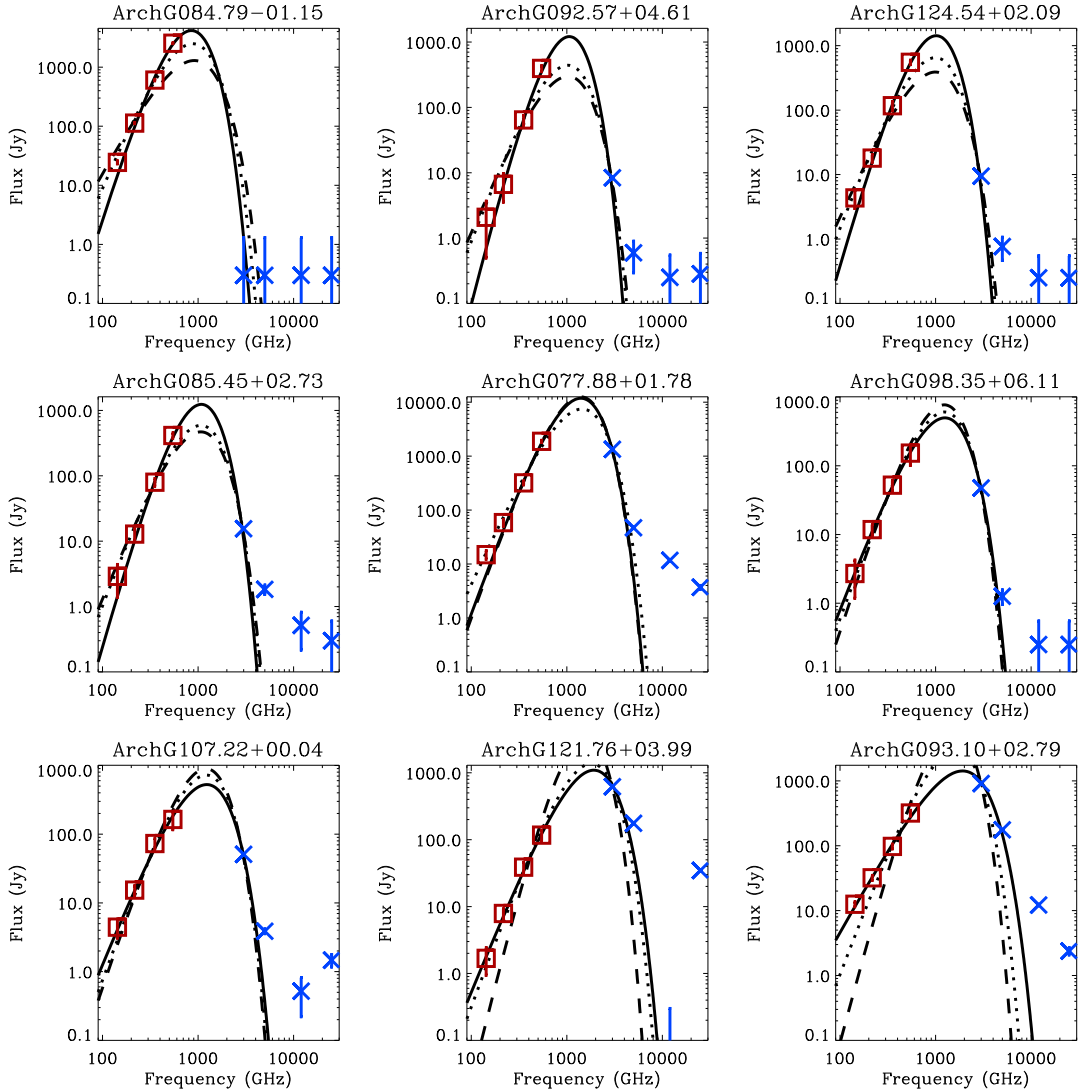}
\end{center}
\caption{Examples of millimetre and far infrared spectra of {\sc
    Archeops} sources. {\sc Archeops} measurements are shown with
  error bars as the red squares. {\sc Iras} measurements (from the
  point-source catalog or the small extended source catalog if
  present, \cite{Beichman1988}) are noted as blue crosses with their
  error bars. A modified blackbody law fit to the {\sc Archeops} and
  the 100~$\mu\mathrm{m}$ {\sc Iras} flux is shown as a solid line.
  The dotted curve is obtained by fixing the emissivity exponent at a
  value of 2. The dashed curve is obtained by fixing the temperature
  at 11~K. Note the steepness of the spectrum of very cold sources in
  the three upper plots (sources ArchG084.79--01.15,
  ArchG092.57+04.61, ArchG124.54+02.09) with respect to the last lower
  plot on the right (source ArchG093.10+02.79).
\label{fig:subspa}}
\end{figure*}

\begin{figure*}[htbp]
\begin{center}
\includegraphics[%
   width=0.7\paperwidth,
   keepaspectratio,
  angle=0]{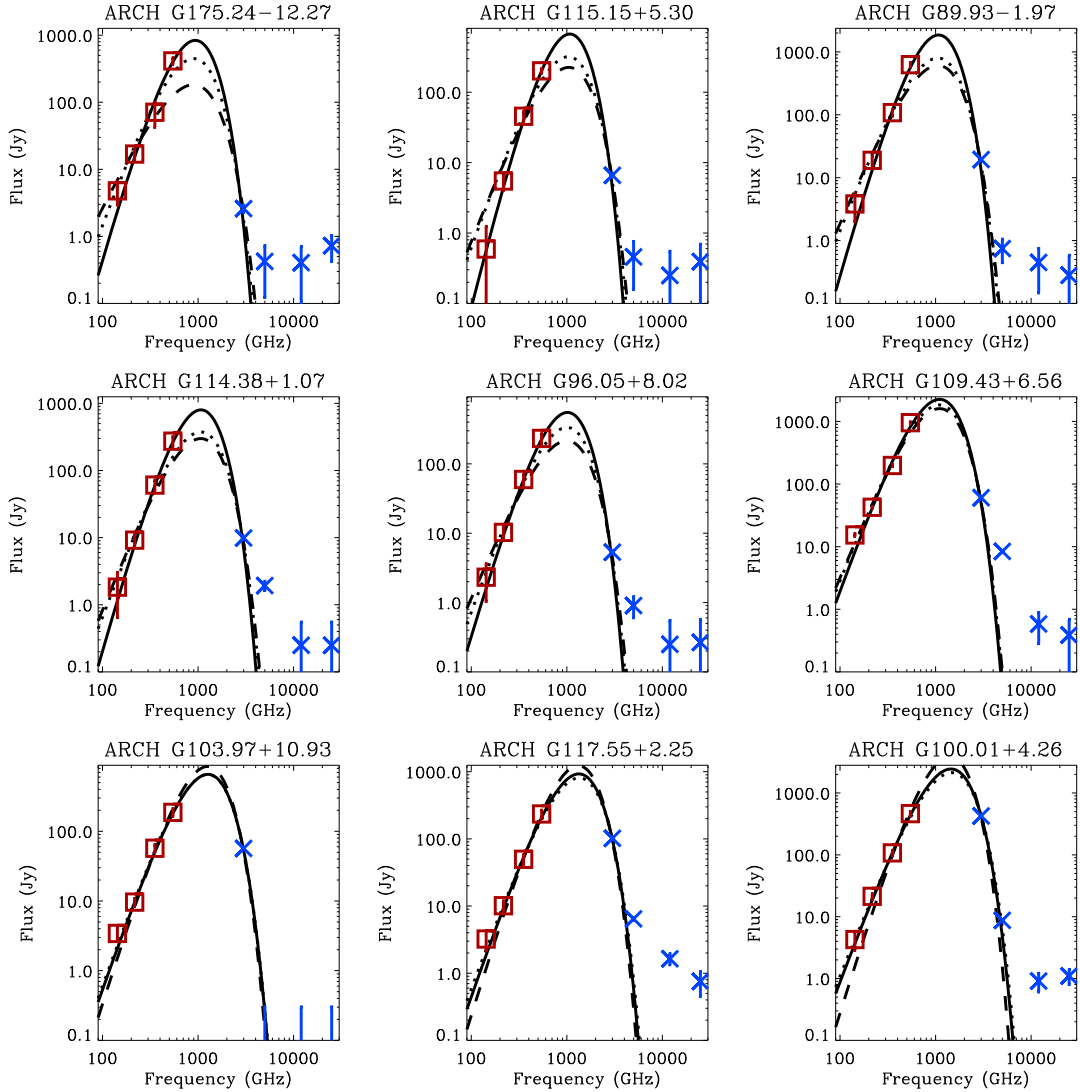}
\end{center}
\caption{ More examples of millimetre and far infrared spectra of {\sc
    Archeops} sources. {\sc Archeops} measurements are shown with
  error bars as the red squares. {\sc Iras} measurements are noted as
  blue crosses. See Fig.~\ref{fig:subspa} for explanations of the
  fitting lines.
  \label{fig:subspb}}
\end{figure*}

Figs.~\ref{fig:subspa} and \ref{fig:subspb} show the millimetre and
infrared spectrum of a sample of the {\sc Archeops} sources along with
the {\sc Iras} measurements.

We have fitted these spectra with a simple modified blackbody law:
\begin{equation}
F_\nu\propto \nu^\beta B_\nu(T)\,\mathrm{,}\label{eq:fnubnu}
\end{equation}
with a single dust component at a temperature $T$ and an emissivity
index of $\beta$.  The {\sc Iras} flux at 100~$\mu\mathrm{m}$
mostly determines the temperature, whereas {\sc Archeops} data points
lead the fit of the emissivity exponent. It is clear that the peak
emission of these sources happens at the THz frequency and will have
to await {\sc Herschel} and {\sc Planck} observations.

The emissivity law exponent $\beta$ and the temperature $T$ quoted in
Table~\ref{assoc} are found by fitting the four {\sc Archeops} fluxes
and the 100~$\mu\mathrm{m}$ {\sc Iras} flux. The fit is generally good
except for 15\% of the sources where the 353~GHz flux shows a small
systematic deficit, or when {\sc Iras} or some {\sc Archeops} band
measurements are missing. The goodness of the fits has been tested
using a $\chi^2$ (quoted in Table~\ref{assoc}) goodness-of-fit
criteria at the 2-$\sigma$ level.  In Table~\ref{assoc} when the fit
is unsatisfactory, we quote the emissivity index $\beta_{11}$ obtained
by fixing the temperature to 11~K in the fitting procedure, as well as
the temperature $T_2$ obtained by fixing the emissivity exponent
$\beta$ at the fiducial value of 2 (only statistical error bars are
quoted).  In the following we will only consider the global
two-parameter ($\beta$ and $T$) fit unless otherwise stated. The
fiducial temperature of $11\,\mathrm{K}$ was chosen as close to an
average of the temperatures found in the two--parameter fits (see also
Fig.~\ref{fig:beta_temp}).

\subsection{Submillimetre point--source number 
counts}\label{subse:numbercount}

\begin{figure}[htp]
\begin{center}
\includegraphics[%
  width=0.6\columnwidth,
  keepaspectratio,
  angle=90]{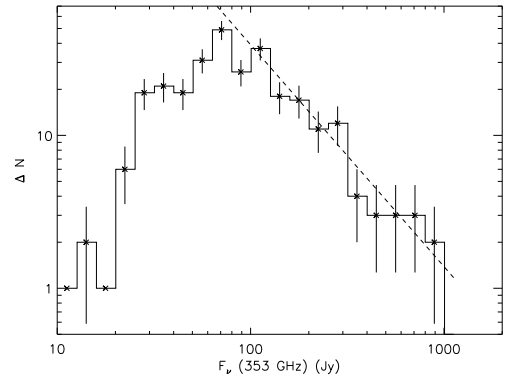}
\end{center}
\caption{\label{fig:lognlogs} Un-normalized number counts of the
  353~GHz {\sc Archeops} sources.  Error bars in each histogram bin
  are assumed to follow a Poisson law.  A best-fit power law with an
  exponent of -1.5$\,\pm 0.2$ is traced as the dashed line.
  Incompleteness starts below 100~Jy flux.}
\end{figure}

Excluding the four brightest sources (Jupiter, Saturn, Crab and Cas
A), the study of the submillimetre spectrum of the {\sc Archeops}
sources indicates that they may constitute a homogenous set of
dust-emission sources.  Therefore, we can compute the number count of
the Archeops sources.  For this purpose, we fiducially choose the
353~GHz data. Fig.~\ref{fig:lognlogs} represents the un-normalized
number counts of the 353~GHz sources as a function of flux. We see a
strong increase of the number counts with decreasing flux down to
about 100~Jy, which is well above the $4\,\sigma$ sensitivity level
for {\sc Archeops}. Then the number counts start decreasing with
decreasing flux. This is mostly due to the sensitivity cutoff of
Archeops at 353~GHz. At the high flux range, the number counts
$F.dN/dF$ is consistent with a power law with an exponent of $-1.5$.

The survey mostly covers the Galactic Plane between the galactic
longitudes of 75 and 198~degrees. We can thus normalize the number
counts to one degree of Galactic longitude. We obtain the following
integral number counts:

\begin{equation}
N(>F_\nu)= (1.0\pm 0.1  \,\mathrm{source/deg})\,
\big(\frac{F_{\nu}(353\,\mathrm{GHz})}{100\,\mathrm{Jy}}\big)^{-1.5\pm
0.2}
\,\mathrm{.}
\end{equation}

The number counts for the other {\sc Archeops} frequencies can be
computed by using a scaling factor of the form
$100\,\mathrm{Jy}\,(\nu/353\,\mathrm{GHz})^4$, where $\nu$ is the
  required frequency .

Assuming a constant latitude width of 10~degrees, the confusion limit,
defined as one source every thirty beams, is $\sim$27~Jy. This is well
below the {\sc Archeops} detection limit. The point--source
contribution to the diffuse background is:

\begin{equation}
B_\nu(>F_\nu)=
0.09\,\mathrm{MJy/sr}\,
\big(\frac{F_{\nu}(353\,\mathrm{GHz})}
{100\,\mathrm{Jy}}\big)^{-0.5}\,\mathrm{,}
\label{eq:intflux}
\end{equation}

which is a negligible fraction of the average brightness in the
Galactic Plane, unless we extrapolate the number counts to sub-Jansky
levels. We can thus conclude that the galactic submillimetre sky is
dominated by diffuse emission and not by point--sources.

\section{Discussion}

Due to the limited angular resolution of {\sc Archeops}, the observed
point-sources are merely clumps of very cold interstellar matter. They
are a sub-population of {\sc Iras} Galactic point-sources or slightly
extended sources which are important for the study of the early stages
of star formation.  We have already noted the photometric
  disagreement between ground--based and the present balloon-borne
  measurements. Wings in shallow column density profile around very
  dense clumps might explain the larger fluxes in {\sc Archeops}
  photometry. Also, in order to prepare the analysis of future high
frequency CMB experiments like the {\sc Planck} satellite mission, it
would be important to be able to simulate these very cold clumps in
the whole sky using their spectral behavior and high frequency surveys
like {\sc Iras}. We now discuss their statistical properties.

\subsection{Submillimetre spectral properties}

\begin{figure}[ht]
\begin{center}
  \includegraphics[width=0.8\columnwidth,
  keepaspectratio]{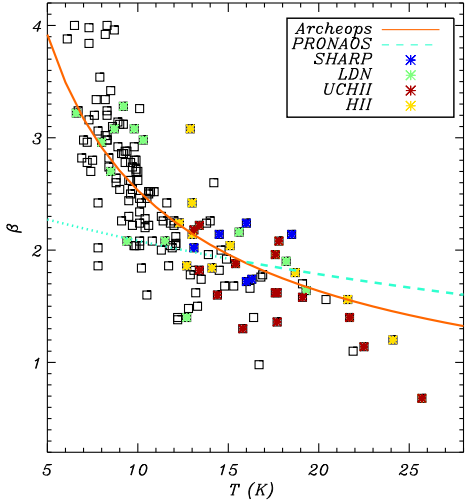}
\end{center}
\caption{Dust emissivity exponent $\beta$ vs. temperature $T$ for the
  {\sc Archeops} sources (black squares).  With red, green and blue
  stars we represent those {\sc Archeops} sources identified as UCHII
  (ultracompact HII), SHARP (Sharpless HII) and LDN (Lynds Dark
  Nebulae) sources respectively. With the orange solid line and light
  blue dashed line we overplot the {\sc Archeops} and {\sc Pronaos}
  $\beta$--$T$ relationships as discussed in the text.
  \label{fig:beta_temp}}
\end{figure}

\begin{figure}[htbp]
\begin{center}
\includegraphics[%
  scale = 0.45,
  angle=0]{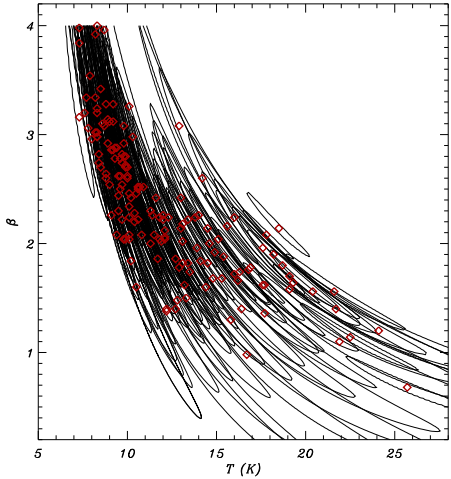}
\end{center}
\caption{\label{fig:betaTlike} Likelihood contour plot in a
  $\beta$--$T$ plot. Around each point source, a contour gives an
  uncertainty equivalent to a 2--$\sigma$ level. Note that the
  correlation between the 2 parameter uncertainties makes it difficult
  to ascertain the true correlation between the parameters.}
\end{figure}

\begin{figure}[htbp]
\begin{center}
\includegraphics[%
  scale = 0.40,
  angle=0]{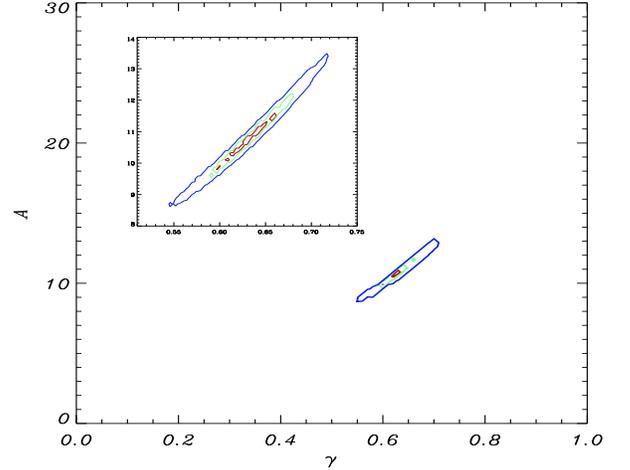}
\end{center}
\caption{\label{fig:AGammalike} Likelihood diagram in the $A$-$\gamma$
  parameter plane (see Eq.~\ref{eq:agamma}) marginalized over $T$.
  Contours correspond to the 0.68, 0.95, 0.9999 C.L. $\gamma = 0$ is
  excluded by the data at more than 0.9999 C.L. indicating a real
  dependence between $T$ and $\beta$. Embedded plot is an expanded
  version of the diagram centered on the maximum of the likelihood.}
\end{figure}

In Fig.~\ref{fig:beta_temp} we trace the emissivity law exponent
$\beta$ as a function of temperature $T$ for the best-fit model to the
spectrum of the {\sc Archeops} sources. We consider only those sources
for which the fit satisfies the $\chi^2$ goodness-of-fit criteria at
95\% C.L.  We find that most of the sources have low temperatures
ranging from 7 to 27~K.  The emissivity exponent increases with
decreasing temperature, going up to 4 for sources below 10~K.

The interpretation of this trend is made difficult by the intrinsic
correlations between $\beta$ and $T$ parameters, as shown in
Fig.~\ref{fig:betaTlike}, where the 1-$\sigma$ confidence level
contour is plotted for each source. We have to assess whether the
observed $ \beta$-$T$ trend can be completely due to the natural
correlation of errors. For this purpose, we parametrize the trend with
a simple power-law:

\begin{equation}
\beta = A T^{-\gamma}\,\mathrm{.}
\label{eq:agamma}
\end{equation}

Fig.~\ref{fig:AGammalike} represents the 1, 2, 3~$\sigma$ confidence
level marginalized over $T$ for the likelihood in the $A$-$\gamma$
plane. A constant $\beta$ value ($\gamma=0$) is clearly excluded at
better than 99.9~\% C.L.

The best fit is given by the empirical law:

\begin{equation}
\beta = (11.5 \pm 3.8) \times T^{-0.66 \pm 0.054}\,\mathrm{,}
\end{equation}

with error bars, after marginalization, at the 0.997 C.L. The standard
$\beta=2$ is obtained for $T=14.1\,\mathrm{K}$. This law is
overplotted (solid line) in Fig.~\ref{fig:beta_temp}. This inverse
$\beta$--$T$ relationship could also be biased by the selection of the
{\sc Iras} counterparts to the {\sc Archeops} sources.  To prove that
this is not the case, we have obtained the {\sc Iras} fluxes at the
position of the Archeops sources by applying the {\sc Archeops}
map-based source extraction algorithm (see Sect.~\ref{psextraction})
to the IRIS maps (\cite{iris_paper}).  These fluxes were then used to
repeat the fit to the spectrum of each of the {\sc Archeops} sources
as above. The new fit leads to the same conclusions.
\cite{dupac_paper} have also claimed an anticorrelation between
$\beta$ and $T$ for dust sources with temperatures in the range
12-20~K, as measured with the {\sc Pronaos} balloon experiment.  For
comparison we trace (with a dashed line, and a dotted line for the
extrapolation to lower temperatures) in Fig.~\ref{fig:beta_temp} the
$\beta$--$T$ relationship they obtained.  {\sc Archeops} and {\sc
  Pronaos} data show a similar behavior, even though the amplitude of
the variations is larger for {\sc Archeops}.  The inverse $\beta$--$T$
relationship is not explained by standard interstellar dust models and
would require us to invoke specific modifications of the optical
properties of the dust, such as the ones produced in amorphous
disordered material, as described by \cite{meny}.  According to such
models, $\beta$ varies not only with temperature but also with
wavelength.  Therefore, the slight differences observed between the
$\beta$--$T$ relationship by the two experiments could be due to {\sc
  Archeops} observing at longer wavelengths (500$\,\mu$m--2~mm) than
{\sc Pronaos} (200--500$\,\mu$m).  Note that laboratory measurements
of dust analogs exist that have revealed an increase of $\beta$ with
wavelengths (see \cite{Boudet2005}), in qualitative agreement with the
observations.  The observed effect is in the opposite direction to the
diffuse interstellar medium colour trend, where $\beta$ is measured to
be low at low temperatures (\cite{Lagache}).  To measure the absolute
value of the emissivity law (which is beyond the scope of this paper)
and scale it to near infrared and visible values, would require an
exhaustive coherent analysis combining the submillimetre and
far--infrared emission properties with the near infrared extinction
properties of these clouds.

\subsection{Submillimetre spectrum compared to far infrared IRAS colours}

\begin{figure*}[htbp]
\begin{center}
\includegraphics[%
  scale = 0.45,
  angle=0]{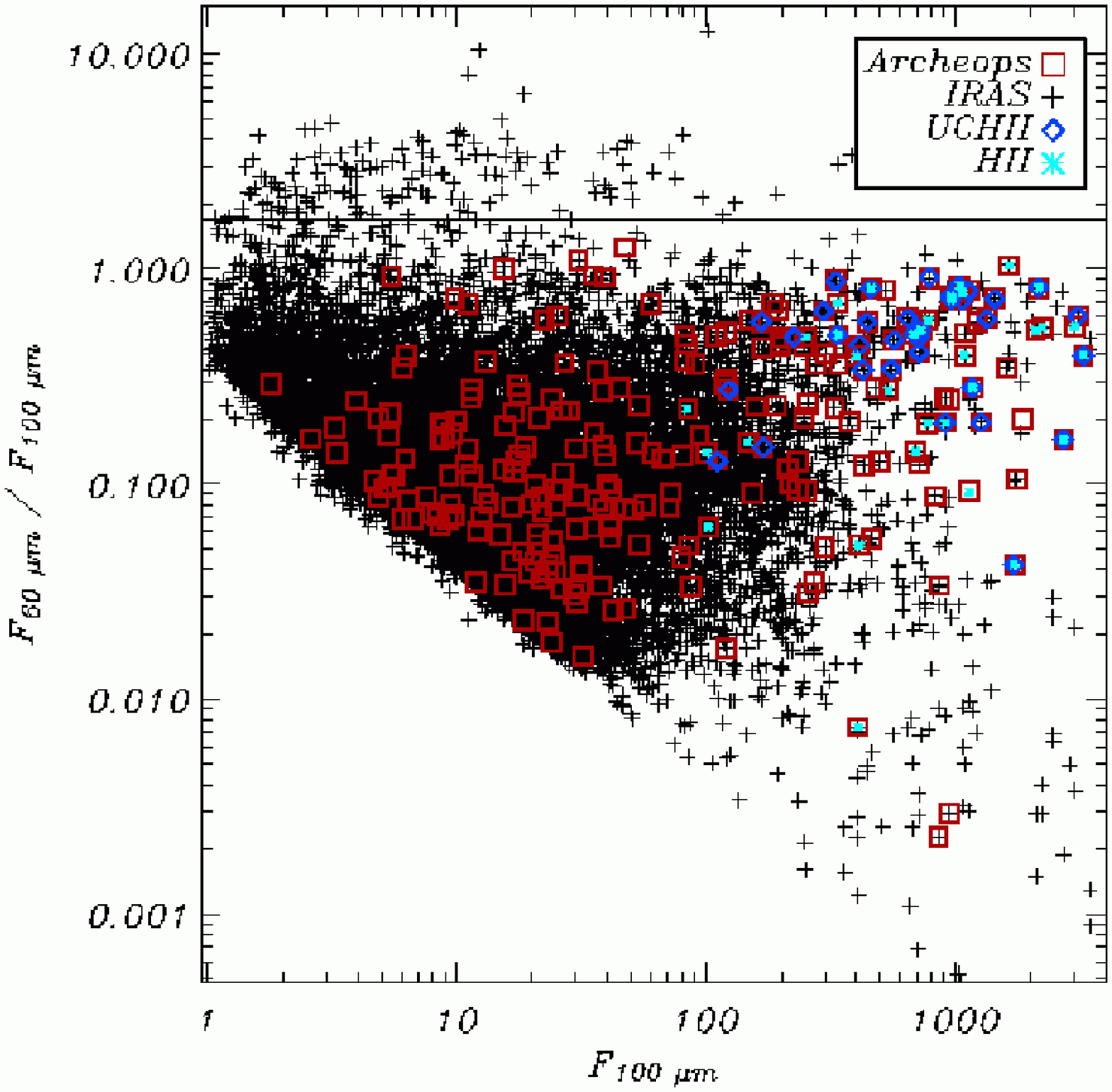}
\includegraphics[%
  scale = 0.45,
  angle=0]{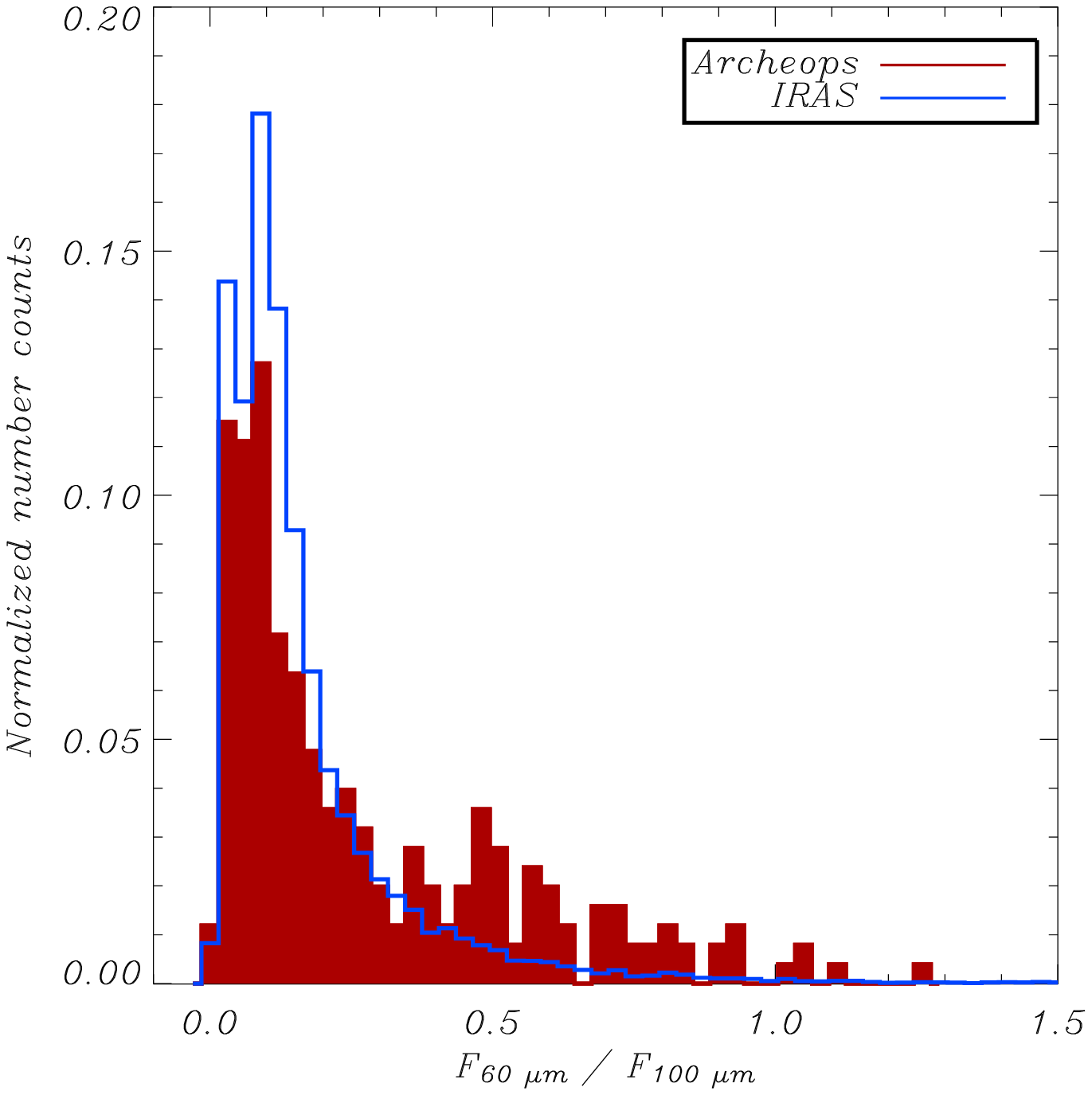}
\end{center}
\caption{\label{fig:irascol} Left plot: Far infrared {\sc Iras} colour
  of point sources as a function of $100\,\mathrm{\mu m}$ {\sc Iras}
  flux. The pluses correspond to the whole {\sc Iras} point-source
  catalog at low galactic latitudes ($|b|<10\,\mathrm{deg.}$) with
  detections at both 60 and $100\,\mu\mathrm{m}$.  The red squares are
  the {\sc Iras} colours of {\sc Archeops} point-sources (when
  detected in {\sc Iras}).  HII regions are concentrated on high
  fluxes and warm colours.  Right plot: Histogram of the far infrared
  {\sc Iras} colour for the whole {\sc Iras} catalogue (blue) of point
  sources at low galactic latitudes ($|b|<10\,\mathrm{deg.}$) with
  detections at both 60 and $100\,\mu\mathrm{m}$ and for the subsample
  of {\sc Archeops} point-sources (red-filled). The warm colour excess
  in the {\sc Archeops} distribution is due to HII regions.}
\end{figure*}

We check whether these very cold clumps present special features in
their far-infrared {\sc Iras} colours that would make them easy to
identify at high frequency.  The left plot of Fig.~\ref{fig:irascol}
shows that the {\sc Iras} far-infrared colours of these clumps broadly
lie in two regions. For large infrared fluxes, the clumps have
relatively warm colours. These sources likely contain HII regions, as
exemplified by the association catalog (Table~\ref{assoc}). On the
other hand, the majority of {\sc Archeops} sources have far-infrared
colours similar to the general galactic point-source value ($\sim
0.2$). More quantitatively, in the right plot of
Fig.~\ref{fig:irascol}, we show the normalized histogram of the far
infrared {\sc Iras} colours for the $|b|\le 10\,\mathrm{deg}$ {\sc
  Iras} point-sources catalogue (light blue) with respect to the
histogram of {\sc Archeops} sources (red--filled).  We observe that
the distribution of the {\sc Archeops} point-sources is similar to
that of the Galactic {\sc Iras} point-sources with an excess of {\sc
  Archeops} sources on the warm {\sc Iras} side.

\begin{figure}[htbp]
\begin{center}
\includegraphics[%
  scale = 0.45,
  angle=0]{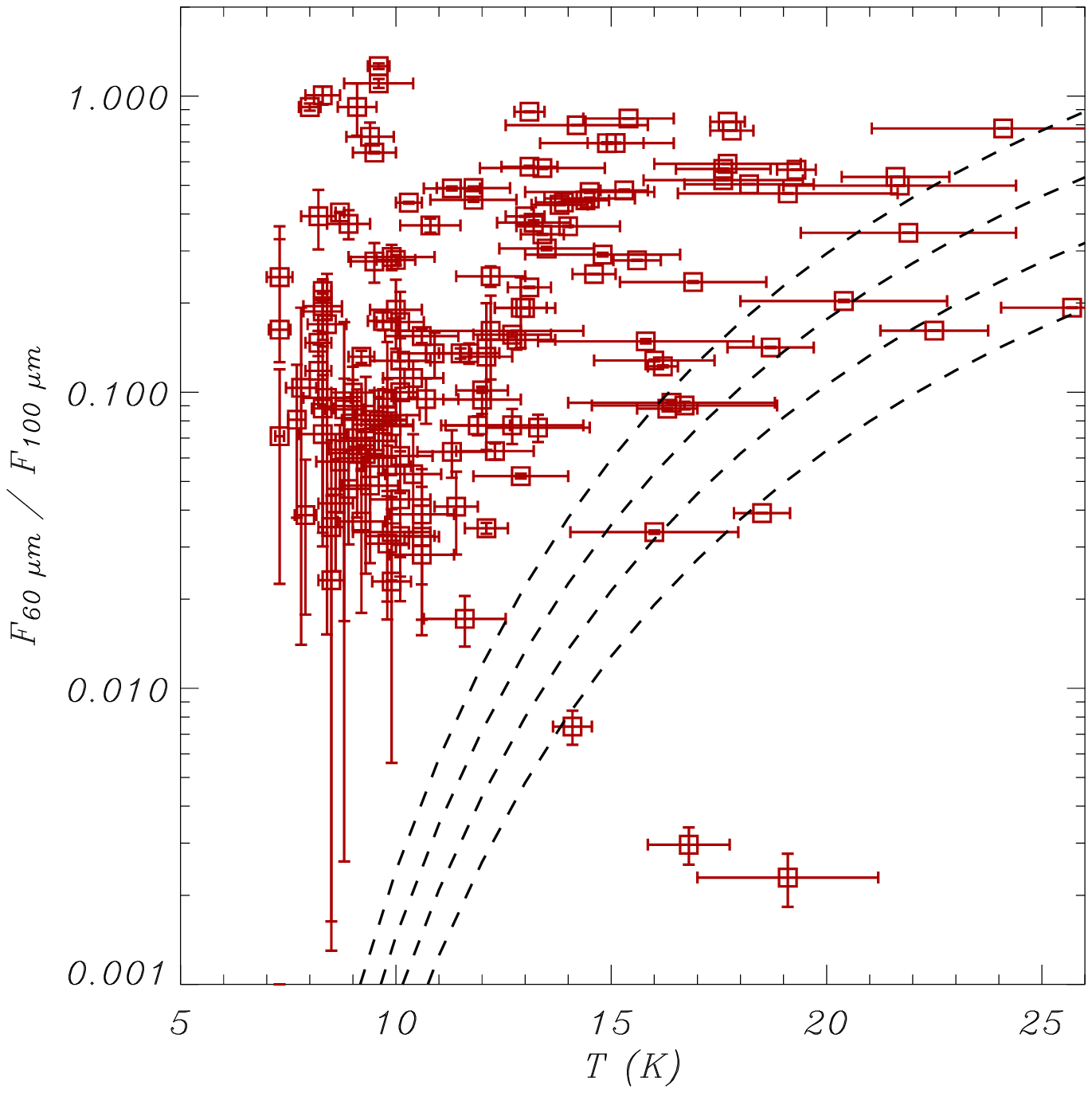}
\includegraphics[%
  scale = 0.45,
  angle=0]{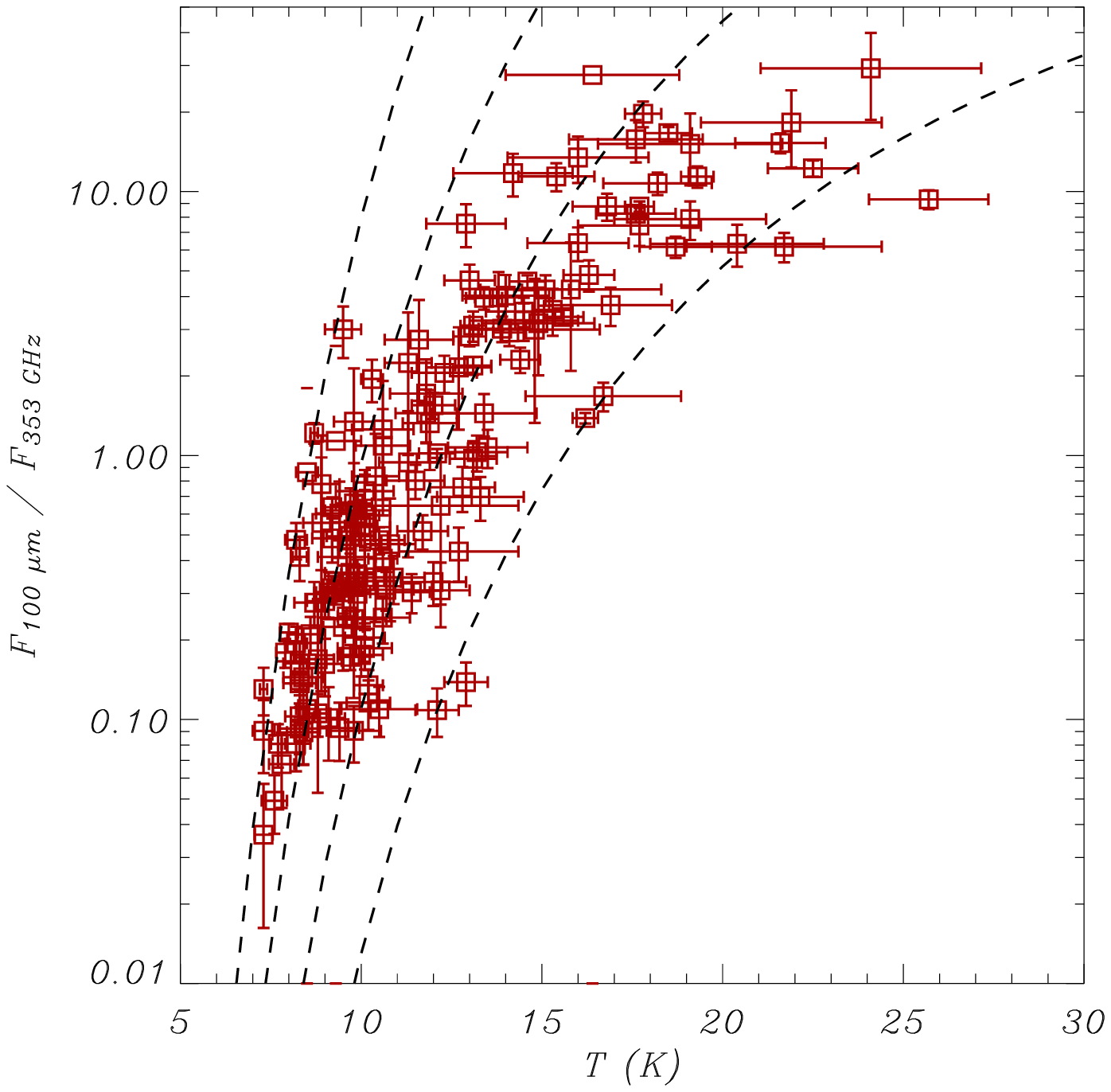}
\end{center}
\caption{\label{fig:irascorr} Upper plot: Far infrared {\sc Iras}
  colour as a function of $T$ for the {\sc Archeops} point sources
  detected at 60 and $100\,\mathrm{\mu m}$. Dashed lines in the two
  plots correspond to single temperature component models with an
  emissivity index of 1, 2, 3, 4, from bottom to top. The two points
  at the lower right of the lines are due to an incorrect association
  with a $60\,\mathrm{\mu m}$ source. Lower plot: Submillimetre colour
  ($100\,\mathrm{\mu m}$ to 353~GHz flux ratio) of {\sc Archeops}
  point sources with a detection at $100\,\mathrm{\mu m}$, as a
  function of $T$ for an acceptable fit of the spectral energy
  distribution.}
\end{figure}

We trace in the upper plot of Fig.~\ref{fig:irascorr}, the far
infrared {\sc Iras} colours for the Archeops sources as a function of
temperature $T$. The far infrared {\sc Iras} colours of the {\sc
  Archeops} sources seem to be uncorrelated with the submillimetre
temperature. For most of the sources, one dust component cannot
simultaneously explain the submillimetre and far--infrared colours. On
the contrary, in the lower plot of Fig.~\ref{fig:irascorr}, a single
dust component is able to explain the submillimetre colours, albeit
with a surprisingly large emissivity index. Therefore the far-infrared
60 to $100\,\mathrm{\mu m}$ ratio is not representative of the
submillimetre dust temperature. Several dust components are needed for
most sources to explain the infrared and submillimetre spectra
together.

From this analysis, we find that the {\sc Archeops} sources show no
distinct features in their far infrared {\sc Iras} colours, except for
the UCHII subpopulation.  This indicates that we cannot easily predict
the fluxes at submillimeter frequencies of known {\sc Iras}
point-sources.  Only $\sim 1.3\,\%$ of the {\sc Iras} sources at low
galactic latitudes ($|b|\le 10^\circ$) are detected by {\sc Archeops}
with a flux $F_\nu(353\,\mathrm{GHz})\ge 50\,\mathrm{Jy}$.

Furthermore, a significant fraction ($\sim 10\%$) of {\sc Archeops}
point-sources are not associated with any {\sc Iras} point-source
making it even more difficult to estimate the point-source properties
in future submillimeter surveys like {\sc Planck} HFI.

\subsection{Ultra compact HII regions.}

Forty {\sc Archeops} sources are associated with ultra compact HII
regions (UCHII, see Sect.~\ref{sect:pointsourcecatalog}). These
regions are of particular physical interest as they are formed by
young stars in the first steps of their evolution. 

In Fig.~\ref{fig:beta_temp} we show as red stars the best-fit
temperature $T$ and dust spectral index $\beta$ for the {\sc Archeops}
sources associated with compact or ultra compact HII regions. We
consider only those sources for which the $T$-$\beta$ fit satisfies
the $\chi^2$ goodness-of-fit criteria at 95\% C.L.: 23 out of 66.
Across the whole temperature range, we observe that these sources
follow the general $\beta$-$T$ relationship described in
Sect.~\ref{ss:submm}: $\beta$ increases for decreasing $T$.  However,
these sources seem to be always hotter than 13~K, likely an effect of
the far--infrared selection. Note that the free--free emission does
not seem to perturb the low--frequency part of the spectrum in these
sources.

We also paid attention to other types of sources like dark Lynds
Nebulas (LDN) and Sharpless HII regions. The best-fit temperature $T$
and dust spectral index $\beta$ for the {\sc Archeops} sources
associated with them are also represented in Fig.~\ref{fig:beta_temp} as
green and blue stars for the LDN and the Sharpless HII regions
respectively.  The {\sc Archeops} point-sources associated with
Sharpless HII regions tend to be hotter than 13~K, similarly to UCHII
regions.  The clouds associated with LDNs seem to be preferentially at
low temperatures ($T < 15$~K) but they follow the general $T$-$\beta$
relationship.

The fact that the submillimeter properties of the HII--associated
clumps are not changed (except that they are hotter) may be
interpreted in the following way: the submillimetre clumps could be
thought of as some precursor stage of the site for massive star
production. Some of those clumps have already started producing
massive stars but those massive stars have not modified the global
properties of the clump enough (except through radiative heating),
perhaps because they have escaped from the cloud, or because the cloud
is much more massive than the stellar cores themselves and hence, the
cloud has not been disturbed yet.

\subsection{Number counts}

In Sect.~\ref{subse:numbercount}, we have seen that a power law
exponent of $1.5\pm 0.2$ gives a fair representation of the integral
number counts. The integrated flux of these clumps 
(see Eq.~\ref{eq:intflux}) is thus dominated
by the faint sources. Because the submillimetre emission is optically
thin, we can directly transform the submillimeter flux $F_\nu$ into a
total clump gas mass using the following formula:

\begin{equation}
M_{\mathrm{gas}}=\frac{F_\nu D^2}{\kappa_\nu B_\nu(T)}=
  1.4\times 10^3\,\mathrm{M}_\odot
\big(\frac{F_\nu(353\mathrm{GHz})}{100\,\mathrm{Jy}}\big)
\big(\frac{D}{1\,\mathrm{kpc}}\big)^2\,\mathrm{,}\label{eq:mgas}
\end{equation} 
where $B_\nu(T)$ is the Planck function and $D$ is the distance to the
source. The last equality is obtained with the dust absorption
coefficient $\kappa_\nu$ taken as $0.0012\,\mathrm{m^{2}.kg^{-1}}$
(\cite{Preibisch1993}) at 353~GHz and an assumed average dust
temperature $T$ of 14~K.

Hence, for a distance range of 200~pc to 2~kpc the detected clumps
have a mass range from $40$ to $40,000\,\mathrm{M}_\odot$,
intermediate between cores and giant molecular clouds. The {\sc
  Archeops} beam encompasses 3.5~pc at a distance of 1~kpc, a
''typical'' intermediate size too.  The likely dispersion in the
distance of the different clumps prevents an interpretation of the
flux number counts in terms of mass distribution and favors
approaching an Euclidian count with an exponent of 1.5~.  For example,
clumps close to Cygnus are likely at a distance of 1.7~kpc
(\cite{Schneider2006}). Other clumps close to Taurus and Perseus
associations are typically at a distance of about $200\pm 60$~pc
(\cite{Dame}).  Optical depth effects are not likely to play a major
role in the fitting of Eq.~\ref{eq:fnubnu} because a typical mass of
$1,000\,\mathrm{M}_\odot$ extending over $1\,\mathrm{pc}$ has an
opacity of 0.01 only at $545\,\mathrm{GHz}$ (using $\tau=4\kappa_\nu
M_{\mathrm{gas}}/(\pi D^2)$).

The satellite {\sc Planck} (\cite{planck}) with a thousand--fold
increase in sensitivity should supersede the present results. In the
mean time, the number counts allow us to firmly establish that there
will be enough bright point-sources to reconstruct and monitor the
{\sc Planck} HFI (\cite{Lamarre}) focal plane geometry. Indeed, each
detector has an instantaneous sensitivity between 1 and 2~Jy per
acquired sample (about 5 milliseconds of integration). The sources
above 100~Jy will thus be detected with an excellent signal to noise
ratio at each crossing. As the scans will drift by typically one
degree per day, hundreds of sources will be available for accurate
astrometry of the HFI detectors during the course of the {\sc Planck}
survey, roughly one every day, and probably more when crossing the
inner Galaxy.

Moreover, a deep unbiased all-sky survey of submillimetre clumps will
be available at the end of the {\sc Planck} survey. With a final HFI
sensitivity of about 10-50~mJy ($1\,\sigma$), by extrapolating {\sc
  Archeops} number counts to the whole galaxy and to fainter fluxes,
one can expect {\sc Planck} to detect tens of thousands of
submillimetre sources, actually down to the confusion limit, in the
Galactic Plane.

\section{Conclusions}

This is a systematic extraction of the point sources from the {\sc
  Archeops} last flight data. These point sources are valuable for the
number counts of galactic sources in the (sub)millimetre domain and
the study of the early star formation processes.  The mean integration
time per source is only of few tenths of seconds.  Nevertheless, this
shallow but wide survey has uncovered many new members of the family
of very cold galactic clumps.

In general the {\sc Archeops} point-sources can be associated with
sources in the {\sc Iras} point--source catalogue or the {\sc Iras}
small extended source catalogue.  Only 30 {\sc Archeops}
point-sources out of a total of 304 remain unidentified.

The spectral energy distribution of the {\sc Archeops} point-sources
is compatible with a modified black body having as parameters the
temperature $T$ and dust spectral index $\beta$. By fitting the
spectrum of the {\sc Archeops} sources to such a model we find that most
{\sc Archeops} point-sources are cold clumps with temperatures in the
range of 7 to 27~K. We also prove that there exists an inverse
relationship between $T$ and $\beta$: $\beta$ increases significantly
with decreasing $T$.

We have found that most of the 302 {\sc Archeops} point-sources have
standard galactic {\sc Iras} far infrared colours, except for 40
identified as ultra compact HII regions, and 26 identified with other
HII regions. Most of the clumps do not seem to be disturbed yet by
internal sources. The far infrared colours are uncorrelated with the
temperature deduced from the submillimetre fit. Only $\sim 1.3\,\%$ of
the {\sc Iras} sources at low galactic latitudes ($|b|\le 10^\circ$)
have a bright submillimetre counterpart ($F_\nu(353\,\mathrm{GHz})\ge
50\,\mathrm{Jy}$).  This means that it is difficult to predict which
{\sc Iras} source will be detected in submillimetre surveys, for
example, in {\sc Planck} simulation of foreground to the CMB
anisotropies.

\begin{acknowledgements}
  The HEALPix package has been used throughout the data
  analysis~(\cite{healpix}). This research has made use of the VizieR
  catalogue access tool, CDS, Strasbourg, France.  We thank the {\sc
    Archeops} collaboration for their efforts throughout the long
  campaigns. This work has been done with the {\sc Planck} {\sc HFI}
  data processing facility.  We thank Bruno Bezard and Raphaël Moreno
  for very fruitful discussions about planets.
\end{acknowledgements}

   \clearpage
   \onecolumn

\begin{flushleft}\begin{tiny}
\begin{longtable}{|r|l|ll|rrr|rrr|rrr|rrr|}\caption{\label{catalog}Archeops
        point-source catalog. The name of the source gives the
        galactic coordinates in degrees.  Fluxes and 1-$\sigma$ errors
        are in Jy, assuming a Rayleigh-Jeans color and for the 4
        frequencies of 143, 217, 353 and 545~GHz. No available
        measurements are quoted as $\sigma = -1$.
        Upper limits are quoted at the 2-$\sigma$ level. The bright sources with an asterisk have positional uncertainties larger than the nominal 4~arcmin. (1-$\sigma$, see Sect.~\ref{psextraction}).}\\

\hline\hline
   N & Name & RA~(2000)  &   Dec~(2000)  &$F_{143}$ & $\sigma_{143}$  &   S/N  &  $F_{217}$ & $\sigma_{217}$ &    S/N   & $F_{353}$ & $\sigma_{353}$  &   S/N  &  $F_{545}$ & $\sigma_{545}$ &    S/N \\
\hline
\endfirsthead
\caption{continued.}\\
\hline\hline
   N & Name & RA~(2000)  &   Dec~(2000)  &$F_{143}$ & $\sigma_{143}$  &   S/N  &  $F_{217}$ & $\sigma_{217}$ &    S/N   & $F_{353}$ & $\sigma_{353}$  &   S/N  &  $F_{545}$ & $\sigma_{545}$ &    S/N \\
\hline
\endhead
\hline
\endfoot
    1 &  ArchG063.36+11.11 &    19 06 19.8& +32 01 15 &   $  <     6.7 $ & $   3.3 $ & $  <       2 $ & $       22.7 $ & $   4.4 $ & $          5 $ & $         66 $ & $    11 $ & $          6 $ & $        201 $ & $    95 $ & $          2 $ \\ 
    2 &  ArchG073.55+02.63 &    20 05 59.7& +36 51 01 &   $  <     6.4 $ & $   3.2 $ & $  <       2 $ & $       24.0 $ & $   4.0 $ & $          6 $ & $        116 $ & $    23 $ & $          5 $ & $          0 $ & $    -1 $ & $          0 $ \\ 
    3 &  ArchG075.75+00.33 &    20 21 40.8& +37 24 44 &   $       50.4 $ & $   7.8 $ & $          6 $ & $      150.4 $ & $  11.9 $ & $         13 $ & $          0 $ & $    -1 $ & $          0 $ & $          0 $ & $    -1 $ & $          0 $ \\ 
    4 &  ArchG075.97+00.69 &    20 20 49.6& +37 47 51 &   $       18.7 $ & $   3.7 $ & $          5 $ & $       23.0 $ & $   9.2 $ & $          2 $ & $          0 $ & $    -1 $ & $          0 $ & $       1940 $ & $    67 $ & $         29 $ \\ 
    5 &  ArchG076.81+02.14 &    20 17 08.4& +39 18 22 &   $       11.4 $ & $   3.2 $ & $          4 $ & $       52.5 $ & $   6.5 $ & $          8 $ & $         88 $ & $    28 $ & $          3 $ & $        584 $ & $    78 $ & $          7 $ \\ 
    6 &  ArchG077.35+01.83 &    20 20 00.2& +39 34 36 &   $       13.9 $ & $   2.0 $ & $          7 $ & $       76.0 $ & $   8.3 $ & $          9 $ & $        266 $ & $    21 $ & $         13 $ & $       1578 $ & $    46 $ & $         34 $ \\ 
    7 &  ArchG077.59+00.62 &    20 25 51.0& +39 05 09 &   $  <     7.0 $ & $   3.5 $ & $  <       2 $ & $       41.5 $ & $   5.7 $ & $          7 $ & $          0 $ & $    -1 $ & $          0 $ & $        554 $ & $    95 $ & $          6 $ \\ 
    8 &  ArchG077.88+01.78* &    20 21 47.5& +39 59 04 &   $       14.9 $ & $   2.7 $ & $          6 $ & $       58.6 $ & $   8.2 $ & $          7 $ & $        319 $ & $    25 $ & $         12 $ & $       1883 $ & $    75 $ & $         25 $ \\
\hline
\end{longtable}%
 \end{tiny}\end{flushleft}


\begin{thebibliography}{}

\bibitem[Baars \etal (1977)]{Baars1977}
Baars J.~W.~M. \etal, 1977, \aap, 61, 99 

\bibitem[Beichman et al. (1988)]{Beichman1988} 
Beichman, C. A.,
  Neugebauer, G., Habing, H. J., Clegg, P. E., \& Chester, T. J., 1988,
  Infrared astronomical satellite (IRAS) catalogs and atlases. Vol. 1:
  Explanatory supplement

\bibitem[Beno\^{\i}t \etal (2002)]{trapani}
Beno\^\i t, A. \etal 2002, Astropart. Phys., 17, 101

\bibitem[Beno\^{\i}t \etal (2003a)]{archpaper}
Beno\^{\i}t A. \etal 2003a, \aap, 399, No. 3, L19 
  
\bibitem[Beno\^{\i}t \etal (2003b)]{archpaper_cospar}
Beno\^{\i}t A. \etal 2003b, \aap, 399, No. 3, L25
  
\bibitem[Beno\^{\i}t \etal (2004)]{archpolar}
Beno{\^ i}t, A., \etal\ 2004, \aap, 424, 571 

\bibitem[Bernard (2004)]{jpbcospar}
Bernard, J. P., 2004, 35th COSPAR Scientific Assembly, 4558 

\bibitem[Boudet \etal 2005]{Boudet2005}
Boudet N., Mutschke H., Nayral C., Jäger C., Bernard J.-Ph., Henning
T., Meny C., 2005, \apj, 633, 272

\bibitem[Chini \etal (1984)]{chini84}
Chini, R., Kreysa, E., Mezger, P. G., \& Gemünd, H.-P., 1984, \aap,
135, L14

\bibitem[Chini \etal (1986a)]{chinia}
Chini, R., Kreysa, E., Mezger, P. G., \& Gemünd, H.-P., 1986, \aap,
154, L8

\bibitem[Chini \etal (1986b)]{chinib}
Chini, R., Kreysa, E., Mezger, P. G., \& Gemünd, H.-P., 1986, \aap,
157, L1


\bibitem[Dame \etal (2001)]{Dame}
Dame, T. M., Hartmann, D., \& Thaddeus, P., 2001, \apj, 547, 792

\bibitem[Dunne \etal (2003)]{Dunne2003}
Dunne, L., Eales, S., Ivison, R., Morgan, H., \& Edmunds, M., 2003,
{\sl Nature}, 424, 285

\bibitem[Dupac \etal (2003)]{dupac_paper}
Dupac, X. \etal, 2003,\aap, 404, L11-L15


\bibitem[Dwek (2004)]{Dwek2004}
Dwek, E., 2004, \apj, 607, 848

\bibitem[Finkbeiner et al (1999)]{finkbeiner}
Finkbeiner, D. P., Davis, M., Schlegel, D. J., 1999, \apj, 524, 867

\bibitem[Goldin \etal (1997)]{goldin:1997}
Goldin, A. B. \etal, 1997, \apjl, 488, L161

\bibitem[Gomez \etal (2005)]{Gomez2005}
Gomez, H. L.,  Dunne, L.; Eales, S. A., Gomez, E. L., \& Edmunds, M. G.
2005, \mnras, 361, 1012

\bibitem[Gorski \etal (2005)]{healpix}
G\'orski, K. M,  Hivon, E., Banday, A. J., \etal, 2005, \apj, 622, 759 
  ({\tt http://healpix.jpl.nasa.gov})

\bibitem[Green 2004]{Green}
Green, D. A., 2004,  Bull. Astr. Soc. India, 2004, 32, 335, updates at
\\({\tt http://www.mrao.cam.ac.uk/surveys/snrs})

\bibitem[Hern\'andez-Monteagudo \etal (2005)]{carlos_sz}
  Hern\'andez-Monteagudo, C., Mac\'{\i}as-P\'erez, J.F., Tristram, M. \&
  Désert, F.--X., 2006, \aap, 449, 41--48

\bibitem[Hinshaw et al 2007]{Hinshaw}
Hinshaw, G., Nolta, M. R., Bennett, C. L., \etal 2007, ApJS, 170, 288


\bibitem[Kurtz, Churchwell \& Wood (1994)]{Kurtz}
Kurtz, Churchwell \& Wood, 1994, \apj, 91, 659

\bibitem[Lagache \etal (1998)]{Lagache}
Lagache, G., Abergel, A., Boulanger, F., \& Puget, J.--L., 1998, \aap, 333, 709L  

\bibitem[Lamarre, \etal 2003]{Lamarre}
Lamarre, J.-M., Puget, J.-L., Piat, M., \etal, 2003, in Proceedings
of the SPIE, IR Space Telescopes and Instruments, J. C. Mather (Ed.),
4850, 730


\bibitem[Mac\'\i as--P\'erez \etal (2007)]{processing} Mac\'\i
  as--P\'erez, J., Lagache, G., Maffei, B., \etal 2007, \aap,
  467, 1313

\bibitem[Mac\'\i as--P\'erez \etal (2008a)]{crab_paper}
Mac\'\i as--P\'erez, J.F., Mayet, F., Désert, F.-X. \& Aumont, J.,
2008a, \aap, companion paper, submitted

\bibitem[Mac\'\i as--P\'erez \etal (2008b)]{casa_paper}
Macias-Perez, J.F., Mayet, F., Désert, F.-X., 2008b, in preparation

\bibitem[Miville-Desch\^{e}nes \& Lagache, 2005]{iris_paper}
Miville-Dech\^{e}nes, M.-A. \& Lagache, G., 2005, \apjs,157, 302-323 

\bibitem[Mény \etal (2007)]{meny}
Meny, C., Gromov, V.,  Boudet, N., \etal, 2007, \aap, 468, 171


\bibitem[Moreno (1998)]{Moreno:1998} 
Moreno, R., 1998, PhD Thesis, Universit\'e Paris VI

\bibitem[Pajot \etal (2006)]{Pajot2006} 
Pajot, F., Stepnik, B., Lamarre, J.-M., \etal, 2006, \aap, 447, 769

\bibitem[Page \etal 2003]{Page2003}
Page, L., Barnes, C., Hinshaw, G.,Spergel, D. N., Weiland, J. L., et
al., 2003, \apjs, 148, 39

\bibitem[Paladini \etal (2003)]{Paladini}
Paladini, R., Burigana, C., Davies, R. D., Maino, D., Bersanelli,
M. \etal, 2003, \aap, 397, 213 

\bibitem[The Planck collaboration, 2005]{planck}
The Planck collaboration. 2005, \\
Available at:\\
{\tt http://www.rssd.esa.int\-/SA/PLANCK/docs/Bluebook-ESA-SCI(2005)1\_V2.pdf}

\bibitem[Ponthieu \etal (2005)]{ponthieu05}
Ponthieu, N., Mac\'\i as--P\'erez, J.~M., Tristram, M., et al, 2005,
\aap, 444, 327

\bibitem[Preibisch \etal, 1993]{Preibisch1993}
Preibisch, Th., Ossenkopf, V., Yorke, H. W., \& Henning, Th., 1993,
\aap, 279, 577




\bibitem[Schneider \etal, 2006]{Schneider2006} 
Schneider, N.,  Bontemps, S., Simon, R., \etal, 2006, \aap, 458, 855

\bibitem[Tristram \etal (2005)]{tristram_cl} 
Tristram, M.,
  Patanchon, G., Mac\'{\i}as-P\'erez, J.F.  \etal 2005, \aap,
  436, 785

\end{thebibliography}
\end{document}